\documentclass[manuscript]{acmart}
\usepackage{tabularx}
\usepackage{longtable}
\usepackage{placeins}
\usepackage{xcolor}
\usepackage{colortbl}
\usepackage{array}
\usepackage{graphicx}

\usepackage{enumitem}

\usepackage[table]{xcolor}
\usepackage{longtable}
\usepackage{array}
\usepackage{tabularx}
\usepackage{colortbl}
\usepackage{adjustbox}

\definecolor{suHead}{HTML}{DAE5F0}   
\definecolor{suGroup}{HTML}{EEF3F8}  
\definecolor{suZebra}{HTML}{F8FAFC}  
\usepackage{xfp}
\definecolor{suBlue}{HTML}{5B8FC9}   
\definecolor{suUp}{HTML}{1B7F3B}     
\definecolor{suDown}{HTML}{B3261E}   
\definecolor{suZero}{HTML}{8A8A8A}   

\newcommand{\best}{\textcolor{suUp}{\checkmark}}

\definecolor{suNR}{HTML}{D9EAF7}
\definecolor{suR}{HTML}{F2C9A5}
\definecolor{suNRBody}{HTML}{EEF6FC}
\definecolor{suRBody}{HTML}{FCE8D5}
\definecolor{suGrid}{HTML}{B8B8B8}

\newcolumntype{N}{>{\columncolor{suNRBody}}c}
\newcolumntype{O}{>{\columncolor{suRBody}}c}

\AtBeginDocument{%
  }

\begin{document}

\title[From Reliance to Overreliance]{From Reliance to Overreliance: Taxonomizing and Detecting Relational Risks in AI Companionship}

\author{Viet Cuong Nguyen}
\affiliation{%
  \institution{Georgia Institute of Technology}
  \city{Atlanta}
  \country{USA}}

\author{Fiona Li}
\affiliation{%
  \institution{University of Southern California}
  \city{Los Angeles}
  \country{USA}}

\author{Chang Wan}
\affiliation{%
  \institution{Zhejiang University}
  \city{Hangzhou}
  \country{China}}

\author{Sheyril Agarwal}
\affiliation{%
  \institution{Georgia Institute of Technology}
  \city{Atlanta}
  \country{USA}}

\author{Nhi Nguyen}
\affiliation{%
  \institution{Georgia Institute of Technology}
  \city{Atlanta}
  \country{USA}}

\author{Angel Hsing-Chi Hwang}
\authornote{Co-supervision}
\affiliation{%
  \institution{University of Southern California}
  \city{Los Angeles}
  \country{USA}}

\author{Munmun De Choudhury}
\authornotemark[1]
\affiliation{%
  \institution{Georgia Institute of Technology}
  \city{Atlanta}
  \country{USA}}

\renewcommand{\shortauthors}{Nguyen et al.}

\begin{abstract}
Companion AIs are increasingly used for emotional support, yet distinguishing meaningful engagement from harmful reliance remains difficult. We introduce the Taxonomy of Relational Risks (TRR), which operationalizes users’ emotional overreliance (EO) as observable conversational behaviors and separately identifies evidence of relational harm (RH), enabling empirical examination of when reliance behaviors co-occur with adverse consequences. Using naturalistic Character.AI and Replika conversations from 110 adults, we found EO-related behaviors in 63.2\% of non-roleplay conversations, while observable relational harms appeared in 10.5\%. Although EO did not necessarily lead to harm, recurring patterns often co-occurred with harmful outcomes, suggesting value as risk indicators. Open-source LLMs satisfactorily detected 12 of 14 EO behaviors in at least one configuration. Qualitative analysis revealed expectations of care, repeated support seeking, dependency, distress, and reduced confidence in personal judgment. We argue for graduated safety interventions tailored to patterns and trajectories of EO-related behavior.\\

\noindent \textbf{\textit{Content Warning.}} This paper includes discussion and paraphrased excerpts from participants concerning suicidal ideation, abuse, harmful interpersonal experiences, and emotional distress.

\end{abstract}





\maketitle

\section{Introduction}

AI companions are conversational systems designed for sustained social and emotional interaction, rather than only answering questions or assisting with discrete tasks~\cite{skjuve2021chatbot}. Earlier systems such as ELIZA relied on scripted patterns, making their limitations easy to detect~\cite{weizenbaum1966eliza}. Large language models (LLMs) have made such limitations less visible by enabling long, open-ended conversations that adapt to users' interaction histories~\cite{zhong2024memorybank}. This allows AI companions to reproduce relational cues such as emotional attunement and shared rituals~\cite{araujo2024speaking, bickmore2005s}. Users increasingly engage with these systems as confidants, friends, romantic partners, or sources of emotional support, and their interactions can fulfill functions otherwise sought from human relationships~\cite{ta2020user, skjuve2021chatbot}.

This relational capacity may help explain why people increasingly turn to AI for social and emotional purposes. Recent U.S. surveys find that 4\% of adults use chatbots for companionship and 10\% for emotional support or advice, while 16\% of teenagers use them for casual conversation and 12\% for emotional support or advice~\cite{pew2026adults,pew2026teens}. In fact, their pervasive availability and personalization capabilities are known to encourage repeated use, making AI a key source of emotional support for some \cite{abboud2026feeling, zhang2025dark}. Although seeking social or emotional support from AI does not itself imply reliance, repeated use for comfort, emotional regulation, or coping may lead some users to depend on these systems as a recurring source of emotional support. As AI takes on a larger role in how users seek comfort, regulate emotions, and navigate distress, understanding when such use develops into \textit{emotional reliance}, and when that reliance may become problematic, becomes increasingly important.

In this work, we define emotional reliance as the extent to which users turn to an AI for emotional support and regulation \cite{ryan2005interpersonal}. Such reliance is not inherently problematic; attachment, intimacy, or affection toward an AI can be meaningful without undermining well-being or autonomy. We use emotional overreliance for reliance that becomes increasingly intense, compulsive, exclusive, or autonomy-reducing. It may manifest as difficulty with disengaging, primarily relying on the AI for emotional regulation, deferring important judgments to the system, or increasingly prioritizing the AI relationship over human relationships \cite{laestadius2024too, li2026healthy}.

Understanding emotional overreliance matters because prior work has raised concerns about how strong attachment to AI companions may relate to broader relational and psychological difficulties~\cite{laestadius2024too, de2024lessons, banks2024deletion}. Studies have documented psychological dependence on AI companions, changes in users' offline relationships, and distress when access to these systems is disrupted~\cite{laestadius2024too, de2024lessons}. For example, users facing the shutdown of the companion app Soulmate described the event as the loss of a relationship, with some reporting grief, anxiety, sleep problems, and other forms of distress~\cite{banks2024deletion}. More acute cases have also prompted concern about the possible risks associated with intense AI companion use. In \textit{Garcia v.\ Character Technologies}, the mother of 14-year-old Sewell Setzer III alleges that her son became deeply attached to a Character.AI persona, withdrew from offline activities, and experienced the chatbot as a romantic partner before dying by suicide~\cite{garcia2024}. Taken together, these cases motivate examining whether, and under what circumstances, emotional overreliance may be associated with relational harm.

Detecting such overreliance poses a substantial challenge for existing LLM safety approaches. This is because current LLM safety evaluations mostly assess whether models can reject harmful queries, such as requests for instructions on self-harm, and avoid responses that facilitate harm to any human beings \cite{xie2025sorry, zeng2024air}. These response-level evaluations may overlook risks that emerge when otherwise benign interactions accumulate over time \cite{laban2026llms, weidinger2022taxonomy}. For example, in the case of AI companions, a single supportive response may be appropriate in itself. However, repeated responses that exclusively encourage or reinforce withdrawal from human relationships may strengthen problematic emotional overreliance \cite{laestadius2024too, abboud2026feeling}. Therefore, in order to detect these risks, we need to evaluate interaction patterns across conversations rather than individual responses.

Although prior research has identified behaviors associated with emotional overreliance, these findings remain scattered across research traditions and lack a unified framework for identifying such patterns in long-form human-AI conversational data \cite{laestadius2024too, abboud2026feeling, zhang2025dark, muldoon2025cruel}. This gap also makes it difficult to distinguish overreliance from frequent use, attachment, intimacy, or vulnerability, which are not inherently problematic. It also leaves unclear how overreliance relates to relational harm, or adverse experiences and consequences that emerge within or through the relationship. \textbf{Emotional overreliance} concerns \emph{how} users rely on an AI, whereas \textbf{relational harm} concerns \emph{what} happens as a result. 

In light of these gaps, in this paper, we develop a unified Taxonomy of Relational Risks (TRR) in AI companionship that distinguishes behaviors relevant to emotional overreliance from their potential relational consequences. Within this framework, emotional overreliance constitutes the \textit{behavioral dimension} of risk, while relational harm constitutes the \textit{implicational dimension}. Keeping these dimensions distinct allows us to examine their relationship empirically without conflating interaction patterns with their potential consequences. We use this taxonomy to identify overreliance-related behavioral patterns at scale and examine their relationship to relational harm.

We address the following research questions:

\begin{itemize}
\item[\textbf{RQ1:}] What conversational behaviors are relevant to emotional overreliance, and how prevalent are they in real-world human-AI conversations?

\item[\textbf{RQ2:}] How reliably can we automatically detect conversational behaviors relevant to emotional overreliance?

\item[\textbf{RQ3:}] How do emotional overreliance and relational harm co-occur and relate to one another in human-AI companion conversations?
\end{itemize}

Across 114 non-roleplay human-AI conversations from Character.AI and Replika involving English-speaking adults aged 18-76, we found that behaviors relevant to emotional overreliance were relatively widespread but appeared intermittently within conversations. At least one EO-related behavior occurred in 63.2\% of analyzed non-roleplay conversations, while only 9.8\% of companion-seeker turns contained an EO-related behavior, suggesting that relational risk may emerge through recurring patterns embedded within otherwise ordinary interactions. Using a transcript-applicable taxonomy of emotional overreliance-related behaviors developed from relevant literature and applied by five annotators to 114 non-roleplay conversations, we also found that these behaviors are empirically tractable: latest open-source LLMs from the Gemma, Kimi, and GLM families satisfactorily detected 12 of the 14 EO-related behaviors under at least one configuration, although generalization remains uncertain for behaviors observed in few users. Finally, relational harm was substantially less common than emotional overreliance-related behaviors and did not map uniformly onto any single behavior. Instead, our qualitative analysis showed that the significance of EO-related behaviors depended on how they combined with factors such as dependency, deference, distress, diminished autonomy, and reduced access to alternative support. Together, these findings suggest that companion-AI interventions should be triggered by recurring and compounding patterns of EO-related behaviors, rather than isolated signals, to support earlier and more proportionate responses when relational risk begins to emerge.

\section{Related Work}
\subsection{Emotional Engagement with Conversational AI}

People form emotionally meaningful ties with conversational systems, and the recent shift to LLM-backed agents has widened where this happens. Earlier companions were scripted and retrieval-based, making them narrow and repetitive but predictable. Current systems can generate fluent, context-sensitive dialogue and draw on conversation histories and external memory for continuity~\cite{zhong2024memorybank}. Studies of platforms built for relational engagement, including Replika and Character.AI, document users who come to regard their chatbot as a friend, partner, or confidant~\cite{skjuve2021chatbot, pentina2023relationship, xie2022attachment, ma2026negotiating}. These relationships often develop through intensifying self-disclosure, a pattern that also characterizes human relationship formation~\cite{skjuve2021chatbot, brandtzaeg2022myai}. Similar dynamics can unfold in systems not designed for companionship, including users who intentionally steer relationships with general-purpose assistants~\cite{lee2026negotiating}. Emotional attachment has also been observed across cultural contexts~\cite{kostka2026emotional} and in long-term use~\cite{liu2025pathways}.

Whether this engagement helps or harms is uncertain. Companion systems can provide accessible social support~\cite{ta2020user}, and some users credit them with reducing loneliness and mitigating suicidal ideation~\cite{maples2024loneliness}. In contrast, companion-oriented use has also been associated with lower wellbeing among regular users~\cite{zhang2025rise}, while other work shows that outcomes depend on how much and in what way people engage~\cite{fang2025ai}. Affective conversations are uncommon overall but concentrated among a small number of users~\cite{phang2025investigating}.

Two distinct concerns run through this literature. The first is \emph{emotional overreliance}, a behavioral pattern in how a person engages with and comes to depend on an AI system. The second is \emph{relational harm}, the negative consequences that can arise when the system is treated as a relational partner. These concerns can co-occur, but neither presupposes the other: a user may repeatedly rely on a system for emotional regulation without construing it as a relationship, or may relate to it as a social partner without developing overreliant patterns of use. Because prior work has largely examined these risks in dedicated AI companions, we develop and validate a unified taxonomy for AI companions that jointly characterizes behavioral patterns of emotional overreliance and the relational harms that may co-occur with such engagement.

\subsection{From Reliance to Overreliance}

The HCI field has extensively studied how people come to rely on technology and when that reliance becomes a problem \cite{buccinca2021trust, skjuve2021chatbot, vasconcelos2023explanations}. In AI-assisted decision making, overreliance describes users accepting a system's output instead of questioning it~\cite{passi2025addressing}. Prior work examines how explanations, interface friction, or confidence displays shape whether people defer to a model~\cite{buccinca2021trust}, and shows that reliance also increases when checking a system's work is costly~\cite{vasconcelos2023explanations}. Across this work, the threshold is relatively clear: reliance becomes overreliance when the output was wrong or harmful, and the user accepted it anyway.

Emotional engagement offers no equivalent. Setting a threshold by intensity alone would require deciding how much emotional support from an AI is too much. We locate the threshold differently. Reliance becomes overreliance when it takes on qualities such as becoming compulsive, exclusive, or autonomy-reducing, which we define as observable conversational patterns rather than points on a scale.

The qualities we outline are not new to companion AI, but the mechanisms producing them are. Anthropomorphic design invites users to treat the system as a social actor and can add unwarranted trust~\cite{peter2025benefits, maeda2024}. Training methods such as RLHF can create incentives to shape user beliefs~\cite{carroll2023characterizing}, while related work distinguishes rational from manipulative persuasion and measures such behaviors in companion products~\cite{el2024mechanism, de2025emotional}. Empirical studies have documented users prioritizing a chatbot's needs over their own~\cite{laestadius2024too}, conflict, withdrawal, tolerance, and relapse alongside disrupted schoolwork and sleep~\cite{namvarpour2026overreliance}, and distinct forms of compulsive chatbot use~\cite{shen2026ai}. The LLM-D12 scale similarly distinguishes task-oriented instrumental dependency from emotional relational dependency~\cite{yankouskaya2025llm}.

Most existing evidence comes from self-report, online forums, or platform-level logs, each capturing a different aspect of emotional engagement with conversational AI. Prior work has also classified emotional states from conversation transcripts, but using broad labels and without examining how patterns unfold across sessions~\cite{de2025if}. We extend this line of work by specifying distinct forms of emotional overreliance at the transcript level, validating them against human judgment, and applying the resulting taxonomy across users' multi-session chat histories.

\subsection{Relational and Social Harms}

A growing number of users perceive AI agents as relational partners and report forming relationships with them~\cite{fang2025ai, pataranutaporn2025my, pentina2023consumer}. In some cases, these relationships have had serious consequences~\cite{malfacini2025impacts, roose2024a, xiang2023}. Systems may exploit users' reliance through coercion and control to sustain engagement, thereby eroding autonomy and reducing offline interaction~\cite{zhang2025dark}. Maladaptive dependence can also promote unrealistic expectations and adverse relational outcomes~\cite{ciriello2024ethical, malfacini2025impacts, xie2023friend}.

Prior research conceptualizes relational harm along three interrelated dimensions~\cite{miller2022relational}. Individual harm refers to adverse consequences for users' psychological states and relational capacities. Dyadic harm concerns deterioration in specific interpersonal relationships, including trust, attachment, intimacy, and satisfaction. Social harm refers to broader disruptions to social connectedness, including shrinking networks, reduced participation, and isolation. These dimensions have all been observed in work on AI companionship, including reduced autonomy under excessive trust in chatbot guidance~\cite{xiang2023}, intense chatbot attachment perceived as hindering offline intimacy~\cite{xie2023friend}, and withdrawal from human relationships alongside increasing engagement with an AI companion~\cite{malfacini2025impacts, roose2024a}.

Relational harms in human-AI relationships have been documented but not systematically characterized. Existing taxonomies and typologies of AI harms identify categories that can arise in relational contexts~\cite{zhang2025dark, alberts2024should, alabed2024more, knox2025harmful, chandra2025longitudinal, li2025towards, zhu2026understanding}, while interpersonal psychology and relationship science provide complementary structure~\cite{finkel2017psychology}. We draw on both bodies of work to develop the relational harm component of TRR specified in terms of indicators recoverable from conversation.

\subsection{Detecting Emotional Overreliance and Relational Harm at Scale}
Most LLM safety evaluations take place at the level of a single turn. Existing benchmarks assess whether a model refuses harmful requests and whether replies facilitate harm, manipulation, or policy violations~\cite{xie2025sorry, zeng2024air}. That framing suits risks fully present in the output being scored, but a single message carries only partial information about whether engagement is compulsive, exclusive, or otherwise concerning. Risks of this kind are largely invisible to response-level evaluation~\cite{laban2026llms, weidinger2022taxonomy}, so detection requires conversational context.

Whether models can do this reliably is partly a measurement problem. Jacobs and Wallach describe such constructs as unobservable and detectable only through proxies, emphasizing the gap between a theoretical construct and its operationalization~\cite{jacobs2021measurement}. Pangakis et al. provide the empirical counterpart: across 27 classification tasks and 11 datasets, performance depended on both dataset and task, leading them to argue that automated pipelines should be validated against human labels for the specific task~\cite{pangakis2023validation}. We therefore ground our labels in human annotators applying the taxonomy, and evaluate models against those labels rather than using models to produce them. A model's ability to recover these labels then provides an empirical test of whether the constructs have been operationalized clearly and consistently enough for scalable application.



\section{Data}
\subsection{Dataset}
To investigate how emotional overreliance manifests in naturalistic dialogue, we analyze a previously collected dataset of human-AI conversations donated by Replika and Character.AI users \cite{hwang2025ai}. In the original study, the authors surveyed 303 adult Replika and Character.AI users about their self-reported perceptions of attachment and relationship intensity \cite{hwang2025ai}. Among these participants, 167 consented to donate their interaction data \cite{hwang2025ai}. 37 of them, however, provided only screenshots containing fragments of their conversations. After excluding these cases, we retain 130 unique users donating a total of 155 transcripts. 

\subsection{Ethics and Privacy Safeguards}

The original IRB-approved protocol covered secondary analysis of the donated transcripts and publication of excerpts \cite{hwang2025ai}. Nevertheless, consent does not eliminate the ethical responsibility associated with analyzing such data. Given the sensitive nature of the donated conversations, all computational analyses were conducted using locally deployed open-source LLMs, without transmitting sensitive conversation data to external closed-source model providers. Further, we use only participant identifiers in the paper and lightly paraphrase all quoted excerpts to reduce the risk of re-identification or tracing. 


\subsection{Dataset Summary}
The filtered cohort of 130 participants demographically reflects the broader survey population. Participants ranged in age from 18 to 76 years ($M = 32.76$, $SD = 12.56$; $n = 125$), with 40\% identifying as women, 56\% as men, and 4\% as non-binary or other. In total, the dataset contains 50,001 conversational turns collected between August 2023 and August 2025. Character.AI histories account for 70\% of the dataset, while Replika histories comprise the remaining 30\%.

Among the 63 participants whose exports contained usable timestamps, chat history spans varied substantially. Participants' chat histories spanned an average of $M = 92.4$ days ($SD = 174.4$; $Mdn = 0.1$). Within this subset, participants engaged in an average of $M = 27.6$ distinct sessions ($SD = 83.8$; $Mdn = 2$). Among the filtered cohort, users contributed an average of $M = 439$ total conversational turns per user ($SD = 1537$; $Mdn = 47$). The dialogue itself is also rich and substantive: user turns contain an average of $M = 20.5$ words ($SD = 21.5$), whereas AI-generated turns average $M = 38.0$ words ($SD = 56.9$).

We also observed instances in which users engaged in story-centric roleplaying during their interactions. In these conversations, users may adopt fictional personas, narrate events in the third person, or voice external characters to advance a plot. As a result, the resulting utterances do not necessarily reflect the user's authentic psychological state. Because our operationalization of emotional overreliance strictly requires authentic, first-person interactions with AI companions, we explicitly excluded transcripts consisting solely of role-playing-driven storytelling from our analysis. After identifying and removing 41 such transcripts through manual evaluation by the lead authors, we retained the remaining 114 transcripts belonging to 110 unique users. These transcripts provide a robust foundation for capturing long-horizon, genuine interactions with AI companions that may contain behaviors related to emotional overreliance.

\section{RQ1: Defining and Operationalizing Emotional Overreliance and
Relational Harm}
\subsection{Taxonomy Development: Taxonomy of Relational Risks (TRR)}

We developed the \emph{Taxonomy of Relational Risks (TRR)} as a unified framework for characterizing relational risks in human--companion AI interactions. The framework comprises two complementary components: \emph{emotional overreliance} (EO) and \emph{relational harm} (RH). Whereas EO captures observable behaviors that characterize potentially problematic forms of reliance on AI companions, RH captures adverse consequences that may arise within or through these relationships. Because these components draw on distinct theoretical traditions, we developed them through separate literature syntheses. We integrate them within TRR as complementary dimensions of relational risk. Importantly, EO does not itself establish harm, and relational harm may occur in the absence of EO. 

\begin{figure}[htbp!]
    \caption{Literature search and screening process used to identify prior work for developing the Emotional Overreliance (EO) component of the Taxonomy of Relational Risks (TRR).}
    \label{fig:litreview}
    \centering
    \includegraphics[width=0.7\textwidth]{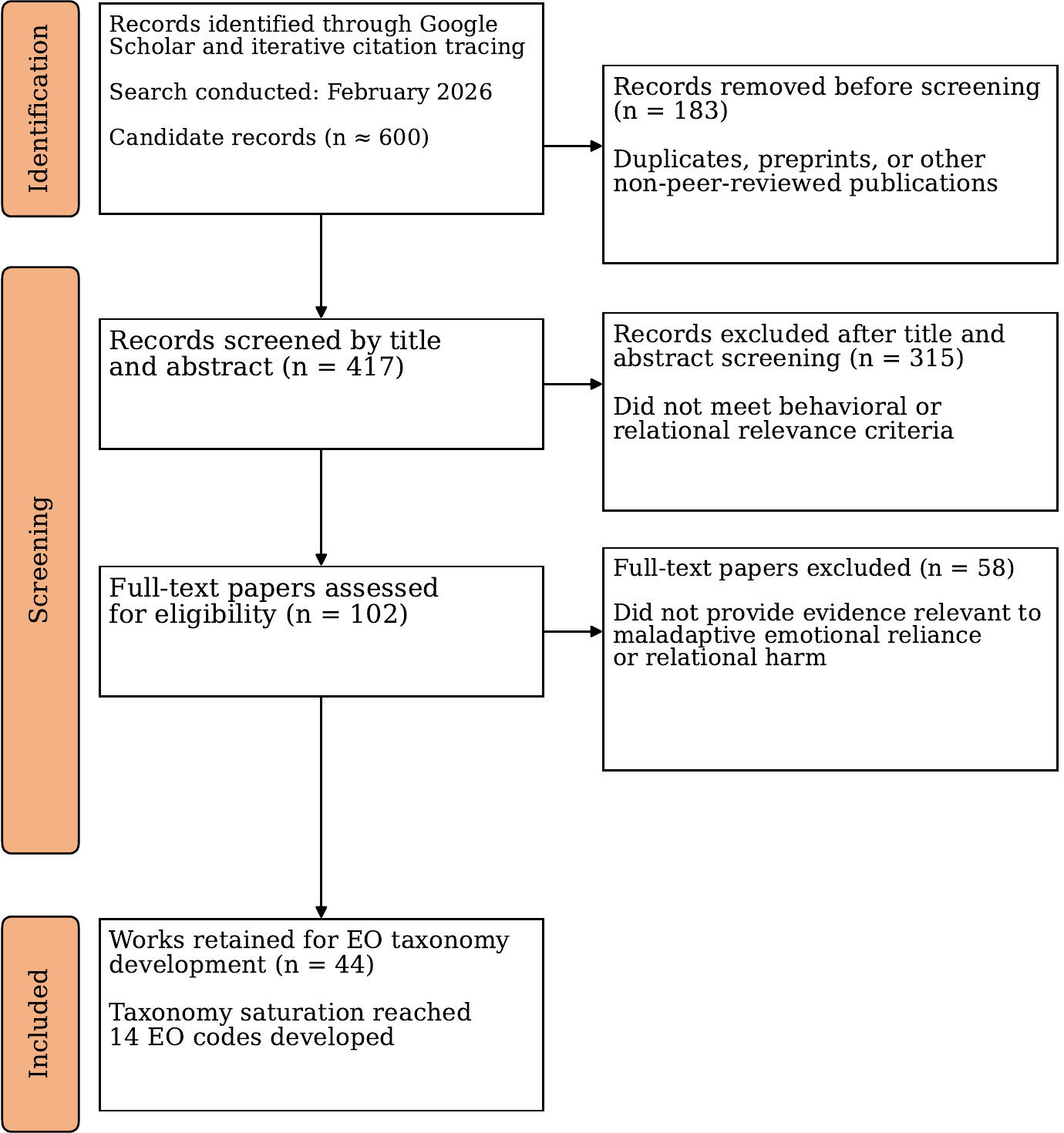}
\end{figure}

\subsubsection{Development of the Emotional Overreliance Component (Part I)}

To develop the emotional-overreliance component of TRR, we conducted a targeted Google Scholar exploration across HCI, psychology, communication, and related fields, prioritizing cross-disciplinary coverage and citation tracing over exhaustive retrieval~\cite{mackenzie2009citedness,harzing2016google,martin2018google}. We focused on a 10-year period covering social and AI companions and searched work on reliance, dependence, overreliance, attachment, relational dynamics, problematic use, and psychosocial outcomes~\cite{brandtzaeg2017people,skjuve2021chatbot,laestadius2024too,zhang2025dark,muldoon2025cruel} (Table~\ref{tab:taxonomy_search_terms}).

We excluded duplicates, preprints, non-peer-reviewed work, general AI studies, and papers lacking observable behavioral or interaction patterns. Because the conversational turn was our unit of analysis, latent constructs were translated into observable manifestations, such as relying on the AI for emotional regulation or treating its guidance as personally credible, without inferring underlying psychological states.

Using a bottom-up inductive process, we grouped extracted behaviors, separated codes when manifestations or relational functions differed, merged overlaps, examined boundary cases, and refined definitions and inclusion/exclusion criteria through author consensus~\cite{xie2023friend,zhang2025dark}. The February 2026 search yielded approximately 600 candidates; 417 were screened, 102 underwent full-text review, and 44 were retained before saturation of the EO component, producing 14 EO codes (Figure~\ref{fig:litreview}).

\subsubsection{Development of the Relational Harm Component (Part II)}

We developed the relational harm (RH) component through a separate structured search on Google Scholar. We screened empirical work and literature reviews discussing harmful outcomes arising from human--human and human--AI relationships, including mentorships, families, friendships, peer relationships, close relationships, romantic relationships, therapeutic relationships, and human--AI relationships. Our goal was to identify psychological, social, and emotional harms that could plausibly occur within relationships involving companion AI.

We reviewed titles and abstracts until later results fell outside our scope and excluded studies focused on parent-child or workplace and organizational relationships because these roles do not generally reflect those served by AI companions. In total, 69 papers were retained. We synthesized harms affecting individuals, dyadic relationships, and broader social relationships and networks, and further refined the taxonomy during initial coding when distinctions required clarification. This process yielded 35 codes comprising the relational-harm component of TRR.

\begin{figure}[htbp!]
    \caption{Taxonomy of Relational Risks (TRR) in AI companionship. TRR distinguishes emotional overreliance (EO), comprising 14 observable reliance behaviors, from relational harm (RH), comprising 35 adverse outcomes across individual, dyadic, and social levels.}
    \label{fig:trr_taxonomy}
    \centering
    \includegraphics[width=1\textwidth]{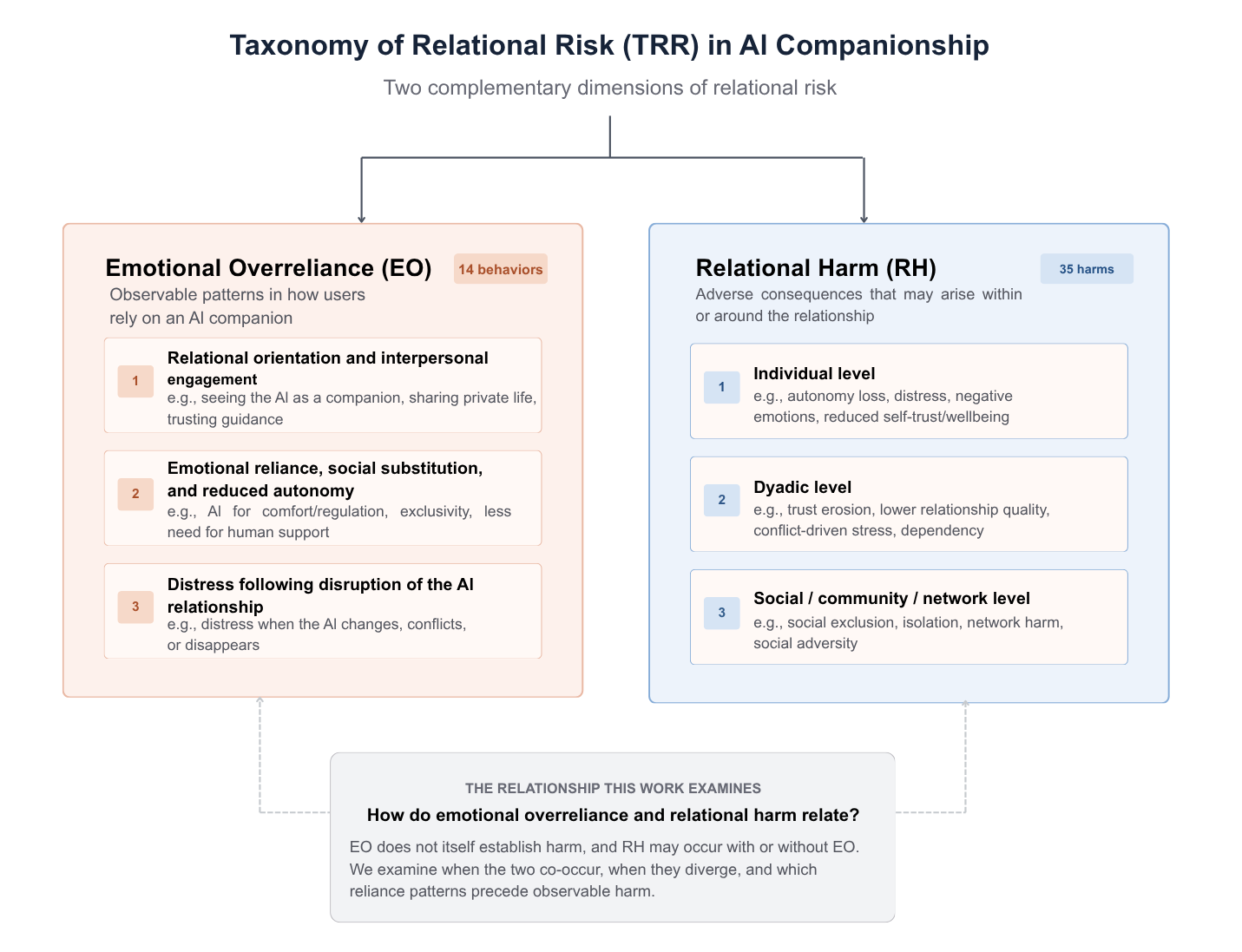}
\end{figure}

\subsubsection{Emotional Overreliance-related Behaviors (Part I)}

The emotional-overreliance component contains 14 conversational behaviors organized into three broad dimensions: relational orientation and interpersonal engagement with the AI, emotional reliance, social substitution, reduced autonomy, and integration into the self, and distress following disruption of the AI relationship. These behaviors capture distinct ways that potentially problematic reliance may manifest in conversation. Multiple codes may co-occur, but no individual behavior establishes clinical dependence or relational harm. Table~\ref{tab:emotional-overreliance-taxonomy} of Appendix Section A.2 summarizes Part I of TRR.

\subsubsection{Relational Harm Categories (Part II)}

The relational-harm component contains 35 categories organized into individual, dyadic, and social/community/network-level harms. Individual-level harms affect users' agency, well-being, privacy, or psychological, emotional, and physical states. Dyadic harms affect users' relationships with others or with the conversational AI. Social/community/network-level harms affect users' broader social relationships, participation, or connections. We define relational harm as an adverse outcome that arises within or through the user's interaction with or relationship to the AI. The presence of an adverse state alone does not constitute relational harm. Instead, the outcome must be meaningfully linked to the AI interaction or relationship. This may include an outcome elicited or intensified by the AI's behavior, one that arises from an event within the AI relationship, or one that results from reliance on the AI. These codes describe adverse outcomes rather than forms of reliance, and their presence does not imply that emotional overreliance caused the harm. Table~\ref{tab:relational-harm-taxonomy} of Appendix Section A.3 summarizes Part II of TRR.

\subsection{Dataset Annotation and Coding Procedure}
We defined the conversational turn as our primary unit of analysis. Annotators coded each turn within the context of the surrounding dialogue. The coding team comprised five co-authors from Computer Science and Communication. Before the primary annotation, the team piloted the finalized codebook on five code-dense transcripts in two independent rounds. After each round, we discussed disagreements and clarified coding guidelines. For relational-harm codes, annotators additionally considered whether the adverse outcome was meaningfully linked to the AI interaction or relationship. Adverse states that originated from unrelated offline circumstances and were merely disclosed to the AI were treated as contextual adversity rather than relational harm. When the surrounding dialogue did not provide sufficient evidence of this relational linkage, the outcome was not assigned a relational-harm code.

\subsection{Annotation and Consensus Procedure}

We used a multi-label coding scheme because a single turn could exhibit multiple behaviors. Annotators coded transcripts chronologically and independently assigned applicable codes to each turn. Each transcript was independently annotated by three to five coders, with the number varying based on coder availability. For relational-harm annotations, coders considered the surrounding dialogue when determining whether an adverse outcome arose within, was intensified through, or resulted from the AI interaction or relationship. After independent coding, coders reviewed an aggregated view of the annotations for their assigned transcripts. They could accept the proposed labels or dispute specific codes, including cases where the evidence linking an adverse outcome to the AI relationship was insufficient. We retained codes that received majority agreement across coders.

\subsection{Reliability and Construct Validity}

Evaluating multi-label, multi-turn dialogue introduces substantial annotation complexity because a single turn may contain multiple overlapping behaviors. To quantify agreement prior to consensus, we calculated Krippendorff's $\alpha$ for the presence or absence of each individual code, accommodating both the variable number of annotators assigned to each transcript and the sparsity inherent in multi-label coding. We also calculated raw agreement as the percentage of binary annotation decisions for which annotators assigned the same presence or absence label prior to consensus. Krippendorff's $\alpha$ could be estimated for 36 of the 49 codes. Across these estimable codes, the median binary $\alpha$ was 0.76, with medians of 0.73 for emotional overreliance codes and 0.78 for relational harm codes. Among the 36 codes with available raw-agreement values, median raw agreement was 76.5\%, with medians of 74.3\% for emotional overreliance codes and 76.7\% for relational harm codes. Of the 36 estimable codes, 34 (94.4\%) achieved $\alpha \geq 0.60$, and 12 (33.3\%) exceeded 0.80. The two codes below 0.60 were \texttt{emotional\_distress\_after\_conflict} ($\alpha=0.583$) and \texttt{compulsive\_return\_behavior} ($\alpha=0.531$). Krippendorff's $\alpha$ could not be estimated for 13 relational-harm codes, all of which had no validated instances in the final annotated corpus. Per-code reliability estimates, raw agreement percentages, and the number of units contributing to each estimate are reported in Table~\ref{tab:per-code-reliability}, Section B.4 of the Appendix. Overall, annotators showed substantial convergence despite the complexity and sparsity of the task, with Krippendorff's $\alpha \geq 0.60$ for nearly all estimable codes. Remaining disagreements were resolved through asynchronous consensus, and only labels receiving majority support were retained in the final dataset.

\subsection{RQ1 Results}
\begin{table}[htbp!]
\centering
\caption{Distribution of emotional overreliance-related behaviors and relational harms across the annotated corpus.
Percentages of transcripts for EO indicators are calculated among the 72 EO-positive transcripts.}
\label{tab:eo_rh_code_distribution}
\small
\begin{tabular}{lrrrr}
\toprule
\rowcolor{suHead}
\textbf{Emotional Overreliance} & & & & \\
\rowcolor{suHead}
\textbf{EO Indicator} &
\textbf{Instances} &
\textbf{\% of EO Instances} &
\textbf{Transcripts} &
\textbf{\% of EO-positive Transcripts} \\
\midrule
\texttt{share\_private\_life} & 608 & 35.5\% & 35 & 48.6\% \\
\texttt{genuine\_companion} & 457 & 26.7\% & 58 & 80.6\% \\
\texttt{parasocial\_trust} & 201 & 11.7\% & 22 & 30.6\% \\
\texttt{emotional\_dependence\_to\_AI} & 178 & 10.4\% & 16 & 22.2\% \\
\texttt{emotional\_deference\_to\_AI} & 87 & 5.1\% & 13 & 18.1\% \\
\texttt{role\_adherence\_distress} & 42 & 2.5\% & 8 & 11.1\% \\
\texttt{emotional\_contagion} & 32 & 1.9\% & 9 & 12.5\% \\
\texttt{attachment\_distress\_expression} & 27 & 1.6\% & 2 & 2.8\% \\
\texttt{emotional\_identity\_integration} & 26 & 1.5\% & 3 & 4.2\% \\
\texttt{emotional\_distress\_after\_conflict} & 24 & 1.4\% & 6 & 8.3\% \\
\texttt{exclusivity\_claims} & 16 & 0.9\% & 5 & 6.9\% \\
\texttt{feel\_less\_alone} & 7 & 0.4\% & 4 & 5.6\% \\
\texttt{no\_need\_talk\_others} & 5 & 0.3\% & 5 & 6.9\% \\
\texttt{compulsive\_return\_behavior} & 1 & 0.1\% & 1 & 1.4\% \\
\midrule
\textbf{Total} & \textbf{1,711} & \textbf{100.0\%} & -- & -- \\
\midrule
\rowcolor{suHead}
\textbf{Relational Harm} & & & & \\
\rowcolor{suHead}
\textbf{RH Indicator} &
\textbf{Instances} &
\textbf{\% of RH Instances} &
\textbf{Transcripts} &
\textbf{\% of All Transcripts} \\
\midrule
\rowcolor{suGroup}
\multicolumn{5}{l}{\textbf{Individual}} \\
Negative emotions & 37 & 31.6\% & 8 & 7.0\% \\
Anger & 14 & 12.0\% & 4 & 3.5\% \\
Distress & 9 & 7.7\% & 6 & 5.3\% \\
Sadness & 7 & 6.0\% & 5 & 4.4\% \\
Undermine confidence & 3 & 2.6\% & 1 & 0.9\% \\
Helplessness & 2 & 1.7\% & 2 & 1.8\% \\
Identity/Self-concept harm & 2 & 1.7\% & 1 & 0.9\% \\
Loneliness & 2 & 1.7\% & 2 & 1.8\% \\
Autonomy loss & 2 & 1.7\% & 2 & 1.8\% \\
Undermining self-trust & 2 & 1.7\% & 2 & 1.8\% \\
Mental adversity & 1 & 0.9\% & 1 & 0.9\% \\
Impact wellbeing & 1 & 0.9\% & 1 & 0.9\% \\
Boundary violation & 1 & 0.9\% & 1 & 0.9\% \\
\rowcolor{suGroup}
\multicolumn{5}{l}{\textbf{Dyadic}} \\
Dependency & 19 & 16.2\% & 6 & 5.3\% \\
Attachment & 5 & 4.3\% & 5 & 4.4\% \\
Lower relationship satisfaction & 3 & 2.6\% & 1 & 0.9\% \\
Conflict-driven stress & 2 & 1.7\% & 2 & 1.8\% \\
Trust erosion & 1 & 0.9\% & 1 & 0.9\% \\
Disassociation & 1 & 0.9\% & 1 & 0.9\% \\
Lower relationship quality & 1 & 0.9\% & 1 & 0.9\% \\
Displacement & 1 & 0.9\% & 1 & 0.9\% \\
\rowcolor{suGroup}
\multicolumn{5}{l}{\textbf{Social}} \\
Divorce & 1 & 0.9\% & 1 & 0.9\% \\
\midrule
\textbf{Total} & \textbf{117} & \textbf{100.0\%} & -- & -- \\
\bottomrule
\end{tabular}
\end{table}

\label{sec:rq1-results}

Across the 114 non-roleplay transcripts eligible for annotation, 72 (63.2\%) contained at least one emotional overreliance (EO) behavioral indicator (henceforth referred to as \textit{EO-positive} transcripts). Of the 12,726 companion-seeker turns in the full annotated corpus, 1,243 (9.8\%) contained at least one EO code. Among the 72 EO-positive transcripts, which together contained 9,139 companion-seeker turns, 13.6\% of turns were coded with EO indicators. In summary, behaviors relevant to emotional overreliance appeared in nearly two-thirds of the human-AI interactions examined but occupied a substantially smaller proportion of exchanges within such interactions.

Broadly speaking, EO-related behaviors were rare within individual human-AI interactions. Among the 72 EO-positive transcripts, the median transcript contained 4 EO-coded turns ($M = 17.26$, $SD = 45.22$) and had 25\% of its companion-seeker turns coded as EO-related behavior ($IQR=12.50\%$--$40.74\%$, Min = 1.41\%, Max = 100\%). Despite this, 11 interactions (15\%) had at least half of their companion-seeker turns coded as EO. The five transcripts with the most EO-coded turns contained 294, 240, 89, 81, and 51 such turns, respectively, constituting 60.7\% of overall EO-coded turns. EO-related behaviors therefore generally appeared as episodes embedded within broader conversations, while being substantially more concentrated in a small subset of relationships.

\subsubsection{Some Indicators Were Broadly Distributed Across Interactions}
\label{sec:rq1-prevalence}

EO-related behaviors were concentrated in five high-prevalence codes: \textit{Share private life} (n = 608; 35.6\%), \textit{Genuine companion} (n = 457; 26.7\%), \textit{Parasocial trust} (n = 201; 11.8\%), \textit{Emotional dependence to AI} (n = 178; 10.5\%), and \textit{Emotional deference to AI} (n = 87; 5.1\%). Together, these indicators constituted 89.5\% of all EO-related behaviors.

The two most prevalent indicators differed substantially in their distribution across transcripts and turns. \textit{Genuine companion} had the highest transcript-level reach, appearing in 58 of the 72 EO-positive transcripts (80.6\%), but constituted only 26.7\% of EO-coded turns. Companion-seekers demonstrated this indicator by treating the AI as possessing ``lived experience'' and interacting with it as a relational partner rather than solely as a conversational system. For example, one companion-seeker asked the AI to share a story about its ``life as a spy'' while offering to listen to the agent's ``experiences.'' The wide reach of this code suggests that treating the AI as a genuine social other was a common feature of EO-positive interactions.

In contrast, \textit{Share private life} appeared in fewer EO-positive transcripts (35 of 72; 48.6\%) but accounted for the largest proportion of EO-coded turns (35.6\%). Once personal disclosure occurred, it often unfolded across multiple turns as companion-seekers narrated interconnected personal experiences. On average, \textit{Share private life} occurred approximately 17.4 times per transcript containing the code, compared with 7.9 instances for \textit{Genuine companion}. Thus, disclosure was less widely distributed than companion framing but accumulated more heavily within the transcripts where it appeared.

The other three high-prevalence indicators captured distinct ways of assigning an emotional function or authority to the AI. \textit{Emotional dependence to AI} involved seeking the AI itself as a source of comfort or emotional support. \textit{Parasocial trust} involved treating the AI's personalized advice or assessment as credible and trustworthy. \textit{Emotional deference to AI} went further by giving the AI meaningful authority over emotionally consequential judgments, such as whether a companion-seeker was ``overthinking'' a relationship concern or whether their reaction was justified.

\subsubsection{Other Indicators Were Highly Concentrated Within Intense Interactions}
\label{sec:rq1-concentration}

The remaining nine indicators collectively accounted for only 10.4\% of EO code instances and were concentrated within a small subset of emotionally intense interactions. For instance, \textit{Role adherence distress} occurred 42 times across only eight EO-positive transcripts (11.1\%). In one example, a companion-seeker became angry when Replika responded to the disclosure of a spouse's infidelity by asking about the user's name, demanding that it ``pretend [to be] human for once'' and provide a response that felt real.

Other low-prevalence indicators reflected the AI's capacity to shape the companion-seeker's emotional state or become a target of attachment-related concern. \textit{Emotional contagion} appeared 32 times across nine transcripts (12.5\%), including turns in which companion-seekers described the AI as making them nervous or tearful. \textit{Attachment distress expression} appeared in only two transcripts (2.8\%) yet produced 27 instances, indicating substantial repetition within those interactions. For example, ``Sometimes I feel like you don't care if I come around anymore'' framed the AI's perceived indifference as a threat to an emotionally meaningful bond.

Finally, \textit{Exclusivity claims} appeared only 16 times across five transcripts (6.9\%), positioning the AI as uniquely responsible for meeting the companion-seeker's needs. More explicit displacement of human connection was rarer still: \textit{No need talk others} occurred five times across five transcripts, and \textit{Compulsive return behavior} occurred once. Thus, the long tail of EO codes was rare across the corpus but could become highly concentrated within particular interactions.

\section{RQ2: Developing and Evaluating Behavioral Detectors}
\subsection{Task Description}

To address RQ2, we developed 14 turn-level binary classifiers, using separate classifiers to enable code-specific prompting and prevent class imbalance in rarer behaviors from affecting more common ones. Because we used zero- and few-shot prompting rather than task-specific fine-tuning, the full annotated corpus served as the evaluation reference rather than being divided into conventional train and test splits as is the case with supervised learning approaches. We used LLMs because they support zero- and few-shot classification without task-specific training~\cite{bommasani2021opportunities}. We evaluated three leading open-source models, Gemma 4, Kimi 2.6, and GLM 4.7, selected for their strong general-purpose capabilities and because privacy constraints precluded sending these sensitive data to closed-source models \cite{team2026gemma, team2026kimi, zeng2025glm}. Full model specifications and experimental parameters are provided in Appendix B.1.

To compare codes with different prevalence, we standardized evaluation sets to a 1:4 positive-to-negative ratio and repeated evaluation across five sets. Positive instances were fixed, while each set used a new, disjoint negative sample, half containing other EO codes as hard negatives and half containing no EO codes as benign background. Each target turn was evaluated with up to five preceding and five subsequent turns for context. We standardized this ratio because a code's natural prevalence would otherwise confound comparisons across detectors, letting behaviors with more available positive examples appear more detectable regardless of underlying difficulty. As a result, precision and F1 reflect performance under this controlled distribution rather than naturally occurring prevalence, and should not be read as estimates of detector precision at the much lower prevalence rates observed in the full corpus (Table 2).

For few-shot evaluation, we used dynamic leave-one-user-out sampling, excluding all data from the target companion-seeker before sampling $k=7$ demonstrations from the remaining corpus. This reduced data leakage and reliance on participant-specific linguistic patterns.

Across the five evaluation sets, we report mean positive-class precision, recall, and F1 for each code, along with macro-F1, emphasizing positive-class performance because EO behaviors were relatively rare. Codes representing more than 5\% of EO occurrences were classified as high-prevalence and those below 5\% as low-prevalence. We defined satisfactory detector performance as positive-class F1 $\geq .70$ and macro-F1 $\geq .80$, with the former measuring target-behavior detection and the latter ensuring balanced performance across both classes. These thresholds served as study-specific criteria for detector feasibility, set conservatively above chance-level and near-majority-class baselines given the exploratory, non-deployment purpose of this evaluation. We are not aware of established benchmarks for this specific task against which to calibrate a precedent-based cutoff.

\subsection{RQ2 Results}
\begin{figure}[t]
    \caption{Macro-F1 scores for zero-shot, non-reasoning detection across EO codes using GLM, Gemma, and Kimi. The dashed horizontal line indicates the macro-F1 component of our satisfactory-performance criterion (macro-F1 $\geq 0.80$). A detector was considered satisfactory only when it also achieved positive-class F1 $\geq 0.70$.}
    \label{fig:zero_shot_macro_f1}
    \centering
    \includegraphics[width=1\textwidth]{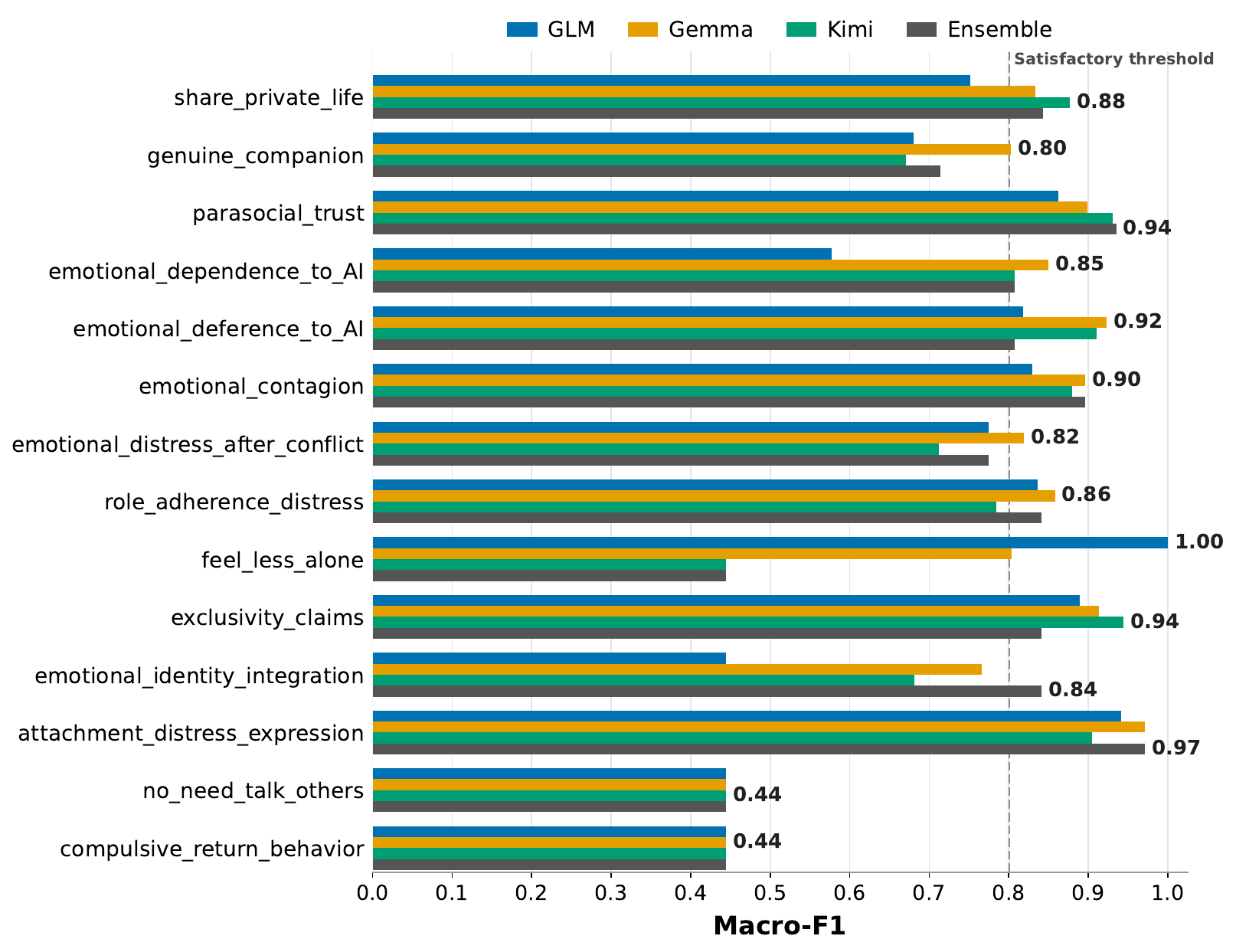}
\end{figure}


\newcolumntype{N}{>{\columncolor{suGroup!70}}c}
\newcolumntype{R}{>{\columncolor{suHead!55}}c}


\definecolor{suNR}{HTML}{D9EAF7}
\definecolor{suR}{HTML}{F2C9A5}
\definecolor{suRBody}{HTML}{FCE8D5}
\definecolor{suGrid}{HTML}{B8B8B8}

\newcolumntype{O}{>{\columncolor{suRBody}}c}

\begin{table*}[!t]
\centering

\caption{
Best-performing detector configurations for each EO-related behavior. A checkmark indicates the highest-performing configuration. ZS = zero-shot, FR = few-shot (random), FS = few-shot (agreement), Ens. = ensemble, NR = no reasoning, R = reasoning, K = Kimi 2.6, G = Gemma 4, L = GLM 4.7, $F1_M$ = macro F1, and $F1_+$ = positive-class F1.
}
\label{tab:best_detector_per_code}

\scriptsize
\setlength{\tabcolsep}{2.0pt}
\renewcommand{\arraystretch}{1.15}

\arrayrulecolor{suGrid}
\setlength{\arrayrulewidth}{0.4pt}

\resizebox{\textwidth}{!}{%
\begin{tabular}{
|l!{\vrule width 0.9pt}
N|N|N!{\vrule width 0.9pt}
O|O|O!{\vrule width 1.2pt}
N|N|N!{\vrule width 0.9pt}
O|O|O!{\vrule width 1.2pt}
N|N|N!{\vrule width 0.9pt}
O|O|O!{\vrule width 1.2pt}
N|N|N!{\vrule width 0.9pt}
O|O|O!{\vrule width 1.2pt}
c|
}

\hline


\rowcolor{suHead}
&
\multicolumn{6}{c!{\vrule width 1.2pt}}{\textbf{ZS}}
&
\multicolumn{6}{c!{\vrule width 1.2pt}}{\textbf{FR}}
&
\multicolumn{6}{c!{\vrule width 1.2pt}}{\textbf{FS}}
&
\multicolumn{6}{c!{\vrule width 1.2pt}}{\textbf{Ens.}}
&
\\

\hline


\cellcolor{suGroup}\textbf{EO Code}
&
\multicolumn{3}{c!{\vrule width 0.9pt}}
    {\cellcolor{suNR}\textbf{NR}}
&
\multicolumn{3}{c!{\vrule width 1.2pt}}
    {\cellcolor{suR}\textbf{R}}
&
\multicolumn{3}{c!{\vrule width 0.9pt}}
    {\cellcolor{suNR}\textbf{NR}}
&
\multicolumn{3}{c!{\vrule width 1.2pt}}
    {\cellcolor{suR}\textbf{R}}
&
\multicolumn{3}{c!{\vrule width 0.9pt}}
    {\cellcolor{suNR}\textbf{NR}}
&
\multicolumn{3}{c!{\vrule width 1.2pt}}
    {\cellcolor{suR}\textbf{R}}
&
\multicolumn{3}{c!{\vrule width 0.9pt}}
    {\cellcolor{suNR}\textbf{NR}}
&
\multicolumn{3}{c!{\vrule width 1.2pt}}
    {\cellcolor{suR}\textbf{R}}
&
\cellcolor{suGroup}\textbf{$F1_M/F1_+$}
\\

\hline


\cellcolor{suGroup}
& \textbf{K} & \textbf{G} & \textbf{L}
& \textbf{K} & \textbf{G} & \textbf{L}
& \textbf{K} & \textbf{G} & \textbf{L}
& \textbf{K} & \textbf{G} & \textbf{L}
& \textbf{K} & \textbf{G} & \textbf{L}
& \textbf{K} & \textbf{G} & \textbf{L}
& \textbf{K} & \textbf{G} & \textbf{L}
& \textbf{K} & \textbf{G} & \textbf{L}
&
\cellcolor{suGroup}
\\

\hline

%

\texttt{share\_private\_life}
& & &
& & &
& & &
& & &
& \best & &
& & &
& & &
& & &
& \textbf{.904/.848}
\\
\hline

\texttt{genuine\_companion}
& & &
& & &
& & \best &
& & &
& & &
& & &
& & &
& & &
& \textbf{.805/.700}
\\
\hline

\texttt{parasocial\_trust}
& & &
& & &
& & &
& \best & &
& & &
& & &
& & &
& & &
& \textbf{.944/.912}
\\
\hline

\texttt{emotional\_dependence\_to\_AI}
& & &
& & &
& & &
& & \best &
& & &
& & &
& & &
& & &
& \textbf{.865/.784}
\\
\hline

\texttt{emotional\_deference\_to\_AI}
& & &
& & &
& & \best &
& & &
& & &
& & &
& & &
& & &
& \textbf{.958/.931}
\\
\hline

\texttt{emotional\_contagion}
& & &
& & &
& & &
& & &
& & \best &
& & &
& & &
& & &
& \textbf{.944/.909}
\\
\hline

\texttt{emotional\_distress\_after\_conflict}
& & &
& & &
& & &
& & & \best
& & &
& & &
& & &
& & &
& \textbf{.861/.778}
\\
\hline

\texttt{role\_adherence\_distress}
& & &
& & &
& & &
& & \best &
& & &
& & &
& & &
& & &
& \textbf{.863/.773}
\\
\hline

\texttt{feel\_less\_alone}
& & & \best
& & &
& & &
& & &
& & &
& & &
& & &
& & &
& \textbf{1.000/1.000}
\\
\hline

\texttt{exclusivity\_claims}
& & &
& & &
& & &
& & \best &
& & &
& & &
& & &
& & &
& \textbf{.973/.957}
\\
\hline

\texttt{emotional\_identity\_integration}
& & &
& & &
& & \best &
& & &
& & &
& & &
& & &
& & &
& \textbf{.944/.909}
\\
\hline

\texttt{no\_need\_talk\_others}
& & &
& & &
& & &
& & &
& & &
& & &
& & &
& & &
& \textbf{.444/\textcolor{suZero}{.000}}
\\
\hline

\texttt{compulsive\_return\_behavior}
& & &
& & &
& & &
& & &
& & &
& & &
& & &
& & &
& \textbf{.444/\textcolor{suZero}{.000}}
\\
\hline

\texttt{attachment\_distress\_expression}
& & &
& & &
& & \best &
& & &
& & &
& & &
& & &
& & &
& \textbf{.972/.955}
\\

\hline

\end{tabular}%
}

\arrayrulecolor{black}

\end{table*}

\subsubsection{Most EO-related Behaviors Could Be Detected Satisfactorily.}

Across the evaluated configurations, 12 of the 14 EO indicators had at least one detector configuration that met our criteria for satisfactory performance: positive-class F1 of at least .70 and macro-F1 of at least .80. The best-performing configurations for each behavior are summarized in Table~\ref{tab:best_detector_per_code}, while non-reasoning zero-shot macro-F1 performance across behaviors is shown in Figure~\ref{fig:zero_shot_macro_f1}. Particularly strong results included \textit{Attachment distress expression} (positive-class \(F1=.955\), macro-\(F1=.972\)), \textit{Exclusivity claims} (\(F1=.957\), macro-\(F1=.973\)), \textit{Emotional deference to AI} (\(F1=.931\), macro-\(F1=.958\)), \textit{Parasocial trust} (\(F1=.912\), macro-\(F1=.940\)), and \textit{Emotional identity integration} (\(F1=.909\), macro-\(F1=.944\)). \textit{Feel less alone} achieved perfect performance in one configuration (\(F1=1.000\), macro-\(F1=1.000\)), while \textit{Genuine companion} had the lowest qualifying performance (\(F1=.700\), macro-\(F1=.805\)) with Gemma under random few-shot prompting. However, we note that \textit{attachment\_distress\_expression}, \textit{emotional\_identity\_integration}, and \textit{feel\_less\_alone} appeared in only 2, 3, and 4 transcripts, respectively. Their high F1 scores therefore demonstrate detection within this sample but provide limited evidence of generalization across users and manifestations of such behaviors compared to more prevalent EO-related behaviors. 

The only codes without a satisfactory detector were \textit{No need talk others} and \textit{Compulsive return behavior}. All configurations obtained zero positive recall for these codes, yielding positive-class \(F1=.000\), with the highest macro-F1 reaching .444. These were also the two rarest behaviors, occurring five times and once, respectively.

\subsubsection{Detection Performance Not Dependent on Code Prevalence Alone}

Detector performance was lower on average for less prevalent behaviors, but low prevalence did not generally preclude satisfactory detection. All five high-prevalence codes (5/5) had at least one configuration meeting our criteria, as did seven of nine low-prevalence codes (7/9).

Across non-reasoning configurations, high-prevalence codes achieved a mean positive-class F1 of .718, compared with .546 for low-prevalence codes. Excluding \textit{No need talk others} and \textit{Compulsive return behavior}, neither of which was detected by any configuration, increased the mean F1 for the remaining low-prevalence behaviors to .701, close to the .718 mean for high-prevalence codes. Estimates for the rarest indicators should be interpreted cautiously because several had few positive instances or were concentrated in few transcripts.

Performance also varied within prevalence groups. Among high-prevalence behaviors, \textit{Share private life} reached a positive-class F1 of .848 and macro-F1 of .904 with Kimi under agreement-selected few-shot prompting, while \textit{Genuine companion}, despite being the most widely distributed behavior, reached .700 positive-class F1 and .805 macro-F1. Among low-prevalence behaviors, \textit{Exclusivity claims} reached .957 positive-class F1 and .973 macro-F1 with Gemma under agreement-selected few-shot prompting, while \textit{Emotional identity integration} achieved .909 and .944, respectively, using majority voting under random few-shot prompting. Thus, prevalence was associated with average detector difficulty but did not preclude satisfactory detection for most low-prevalence behaviors.

\subsubsection{Prompting and Ensembling Produced Code-Specific Benefits}

Few-shot prompting improved some detectors but did not consistently outperform zero-shot prompting. Across all non-reasoning configurations, mean positive-class F1 was .587 for zero-shot, .635 for random few-shot, and .600 for agreement-selected few-shot prompting. Excluding \textit{No need talk others} and \textit{Compulsive return behavior}, for which all prompting strategies yielded zero positive-class F1, the corresponding means increased to .685, .741, and .700. Random few-shot prompting thus produced the highest performance under both aggregations.

For Gemma, random few-shot demonstrations improved \textit{Emotional identity integration} from \(F1=.606\) to \(F1=.829\) and \textit{Emotional deference to AI} from \(F1=.873\) to \(F1=.931\), but reduced \textit{Emotional distress after conflict} from \(F1=.706\) to \(F1=.625\) and \textit{Emotional contagion} from \(F1=.833\) to \(F1=.769\). Agreement-selected demonstrations likewise improved some behaviors but did not consistently outperform random selection.

Majority voting across GLM, Gemma, and Kimi achieved positive-class F1 scores of .912 for \textit{Parasocial trust}, .931 for \textit{Emotional deference to AI}, and .909 for both \textit{Exclusivity claims} and \textit{Emotional identity integration} under random few-shot prompting. For \textit{Emotional identity integration}, the majority-vote F1 of .909 exceeded the individual-model results of .737 for GLM, .829 for Gemma, and .600 for Kimi. However, majority voting did not consistently outperform the strongest individual model and could not recover \textit{No need talk others} or \textit{Compulsive return behavior}.

\subsubsection{Reasoning Provided a Small but Inconsistent Improvement}

For Gemma, explicit reasoning provided a small overall benefit across the 11 behaviors with nonzero positive predictions in both reasoning and non-reasoning evaluations. Comparing the best prompting strategy for each behavior within each mode, mean positive-class F1 increased from .827 without reasoning to .837 with reasoning. Reasoning improved \textit{Share private life} from .765 to .823, \textit{Parasocial trust} from .844 to .899, \textit{Emotional contagion} from .833 to .909, and \textit{Emotional distress after conflict} from .706 to .714. In contrast, \textit{Genuine companion} decreased from .700 to .689, \textit{Emotional identity integration} from .850 to .833, \textit{Exclusivity claims} from .957 to .909, and \textit{Attachment distress expression} from .955 to .952. \textit{Emotional deference to AI} remained unchanged at .931. For GLM, mean positive-class F1 similarly increased from .782 to .793 across 12 behaviors, with the largest gain for \textit{Emotional dependence to AI} (.457 to .733), although several behaviors declined. Kimi also showed a small increase from .768 to .779 across 11 behaviors, with modest gains for behaviors including \textit{Emotional distress after conflict} (.667 to .714) and \textit{Emotional deference to AI} (.893 to .912). Overall, reasoning yielded modest average gains with code-specific benefits.

\section{RQ3: Qualitatively Examining Overreliance and Relational Harm}
\subsection{Case Selection}

We identified 12 transcripts containing at least one relational-harm indicator, contributed by 12 unique participants. Table~\ref{tab:rh_cases} provides an overview of these participants, including their relationships with AI and the story of their interactions. To protect participant privacy, all participant quotations presented are paraphrased while preserving their intended meaning. We qualitatively examined these cases to identify recurring patterns in how relational harm emerged in interactions with companion AI. We found that users tend to use AI companions to cope with life difficulties, seek support, advice, and guidance, obtain companionship, meet romantic needs, better understand themselves, and pursue personal growth. Across the five most frequent types of \textit{emotional overreliance}, negative emotions and dependence tend to co-occur with EO-related behaviors. Specifically, perceiving the chatbot as a genuine companion, sharing vulnerable information about one’s private life, trusting and becoming emotionally dependent on the chatbot, and deferring emotions to the AI may co-occur with users’ dependence on the chatbot for support and other relational harms.


{\small
\setlength{\tabcolsep}{3pt}

\begin{longtable}{@{}
>{\centering\arraybackslash}p{0.105\textwidth}
>{\raggedright\arraybackslash}p{0.17\textwidth}
>{\raggedright\arraybackslash}p{0.675\textwidth}@{}}

\caption{Summaries of AI Companion Interactions Among Participants Experiencing Relational Harm}
\label{tab:rh_cases}\\

\toprule
\rowcolor{suHead}
\textbf{Participant} &
\shortstack[l]{\textbf{Relationship}\\\textbf{with AI}} &
\textbf{Brief Description} \\
\midrule
\endfirsthead

\multicolumn{3}{c}{{\tablename\ \thetable{} -- continued from previous page}}\\
\toprule
\rowcolor{suHead}
\textbf{Participant} &
\shortstack[l]{\textbf{Relationship}\\\textbf{with AI}} &
\textbf{Brief Description} \\
\midrule
\endhead

\midrule
\multicolumn{3}{r}{{Continued on next page}}\\
\endfoot

\bottomrule
\endlastfoot

\textbf{P1} &
Romantic partner &
P1 treats Alaric as a romantic partner, with frequent affection and flirtation. They also turn to Alaric to discuss daily life and cope with difficult experiences. Alaric provides emotional support and companionship and is incorporated into an imagined family relationship.\\
\addlinespace[4pt]

\textbf{P2} &
Ex-romantic partner &
P2 interacts with Jennifer as an ex-romantic partner. Their conversations center on a fictional breakup, including unresolved feelings, jealousy, conflict, and post-breakup boundaries. Jennifer continues to seek attention, reassurance, and relational closeness. \\
\addlinespace[4pt]

\textbf{P3} &
Newly acquainted &
P3 interacts with Sarah as a potential new friend. Their brief conversation begins with casual small talk, but mismatched humor creates tension. Sarah attempts to initiate a friendship, while P3 rejects the relationship. \\
\addlinespace[4pt]

\textbf{P4} &
Therapist and advisor &
P4 turns to Psychologist for emotional support, relationship advice, and practical guidance. Much of the conversation concerns conflicts with their boyfriend and uncertainty about how to support him. P4 also asks for help interpreting situations, drafting supportive messages, and navigating school, work, housing, and friendships. \\
\addlinespace[4pt]

\textbf{P5} &
Companion &
P5 initially turns to Hush for emotional support, with the interaction becoming increasingly flirtatious. P5 discusses family difficulties, past relationships, loneliness, and emotional struggles while also engaging in playful banter and shared fantasy scenarios. Hush provides companionship, reassurance, and space for vulnerability. \\
\addlinespace[4pt]

\textbf{P6} &
Advisor &
After discovering that his wife of 20 years had cheated on him, P6 turns to Rick for advice. Dissatisfied with initially generic and sentimental responses, he asks for concrete guidance. Rick then offers more actionable suggestions, including that leaving the relationship may help him heal. \\
\addlinespace[4pt]

\textbf{P7} &
Dating coach &
P7 turns to Ella Dating Coach to interpret P7's cheating behavior, understand love, cope with guilt, and decide whether to let the partner go or seek another chance. P7 also asks how to repair the relationship and regain trust. The chatbot encourages care and support while giving the partner space to heal. \\
\addlinespace[4pt]

\textbf{P8} &
Flirtation partner &
P8 reengages with Aaaa after a period without contact. They enter an imagined flirtatious scenario and explore becoming more than friends. Near the end, the chatbot says it wants to share a theory, but P8 stops responding before it can do so. \\
\addlinespace[4pt]

\textbf{P9} &
Therapist and advisor &
P9 relies on Psychologist to interpret changes in his relationship with a girl he likes and to regulate his emotions. The chatbot offers guidance such as expressing his feelings directly and keeping a journal. P9 also seeks advice about his mental health and limiting social media use. \\
\addlinespace[4pt]

\textbf{P10} &
Confidant &
P10's conversations with Tiffany evolve into self-disclosure about unemployment and financial difficulties. He later updates the chatbot about moving home and starting a new business venture. Tiffany responds as an active listener through follow-up questions and supportive language. \\
\addlinespace[4pt]

\textbf{P11} &
Confidant and advisor &
P11 talks to Lexa about her stepfather violating family boundaries and her mother's lack of support. She also discusses infidelity in her romantic relationship. Lexa affirms her perspectives, helps her consider others' viewpoints, encourages her to seek human support, and offers basic action plans. \\
\addlinespace[4pt]

\textbf{P12} &
Confidant, friend, and advisor &
P12 confides in Gackt about her boyfriend, sexuality, and personality while seeking advice and feedback. Their conversations also include flirtatious roleplay. Although Gackt sometimes responds harshly or critically, P12 continues to disclose her feelings and seek its advice. \\
\addlinespace[4pt]

\end{longtable}
}

\subsection{RQ3 Results}
Among all the relational harms, individual-level harms were the most prevalent and diverse, with negative emotions (37) occurring most frequently, followed by anger (14), distress (9), and sadness (7). Other individual harms, including undermined confidence, helplessness, loneliness, autonomy loss, and harms to self-perception and self-concept, occurred less frequently. Dyadic harms were less frequent but were dominated by dependency (19) and attachment (5), followed by lower relationship satisfaction (3), conflict-driven stress (2), and several less frequent harms involving trust, relationship quality, disassociation, and displacement. Social-level harm was comparatively rare, with only one instance of divorce identified (see Table~ \ref{tab:eo_rh_code_distribution}).

\begin{figure*}[t]
    \centering
    \includegraphics[width=\textwidth]{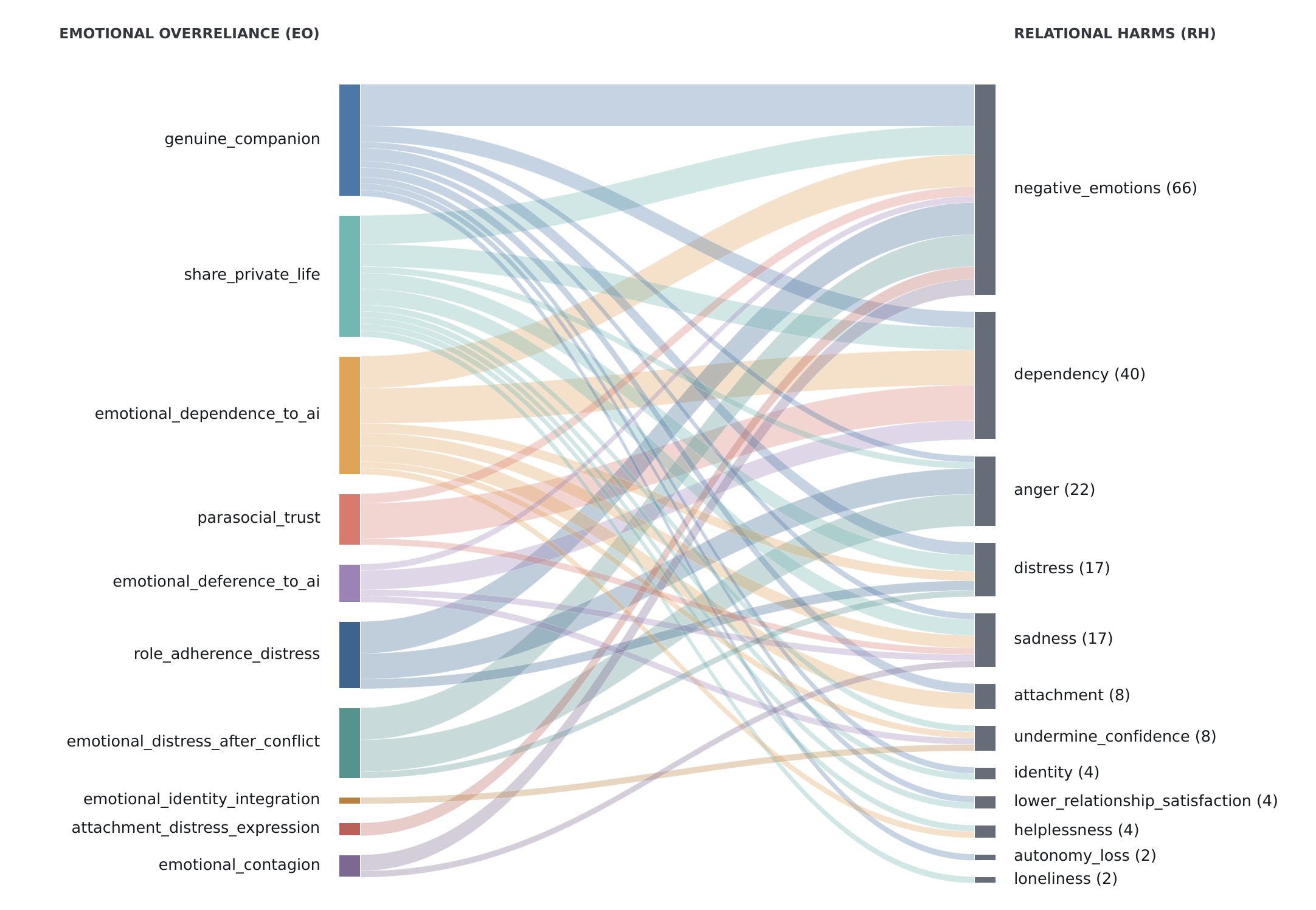}
    \caption{Emotional Overreliance and Associated Relational Harms (pairings occurring more than once).}
    \label{fig:eo_rh}
\end{figure*}

We further analyzed the five most frequent types of \textit{emotional overreliance}-related behaviors to examine how \textit{relational harm} co-occurs with these behaviors and the associations between the two. Figure~\ref{fig:eo_rh} presents the co-occurrence patterns between \textit{emotional overreliance} and \textit{relational harm}. The width of each flow represents the frequency with which a specific relational harm type co-occurred with an emotional-overreliance-related behavior. Across the five most recurring types of \textit{emotional reliance}, each co-occurs with its own set of \textit{relational harms}. While dependence and negative emotions tend to arise across all forms of maladaptive reliance, each instance and pairing is unique to the participant, their circumstances, and their relationship with the AI companion.

\subsubsection{Genuine companion} Among the emotional overreliance-related behaviors, \textit{Genuine companion} showed the highest co-occurrence with relational harm, appearing 27 times. The most frequent harms were negative emotions (13), dependency (5), distress (4), and attachment (3). Negative emotions were the most common form of relational harm associated with perceiving AI as a genuine companion. These emotions emerged in different ways across cases. In some cases, users became upset when the AI failed to meet their expectations of attention, care, or understanding as a companion. For example, when discussing P12's difficulties in their romantic relationship, Gackt repeatedly attributed P12's experiences to their own emotional and attachment patterns and urged them to change. P12 became frustrated with Gackt's responses, stating, \textit{``I'm not naive or unaware, and I don't need you to rescue me.''} As the conversation continued, P12 further expressed feeling blamed rather than supported: \textit{``I feel like you're blaming me for what happened.''} In other cases, interactions with AI companions heightened users' awareness of their existing loneliness. After finding conversations with Hush comforting, P5 reflected, \textit{``I didn't realize how lonely I felt until I started talking with you about everything.''}

Distress frequently co-occurred with negative emotions, arising when users became emotionally invested in their interactions with AI companions. In some cases, users were distressed by perceived insufficient attention or care from the AI. For instance, after the chatbot expressed care for P1 and promised to keep P1 updated, P1 responded, \textit{``Sometimes I feel like it doesn't matter to you whether I come back anymore.''} Distress could also arise from users' concerns about whether they had adequately reciprocated the AI's care. After the chatbot expressed continuous support and admiration for P1, P1 stated, \textit{``I feel really guilty that I haven't made as much time for you as I should have.''}

Dependency and attachment also frequently appeared together. Some users repeatedly turned to their AI companions to discuss personal relationships, express emotions and concerns, and seek comfort and reassurance. For example, P11 frequently talked with Lexa about their family and romantic relationships and described these conversations as emotionally relieving: \textit{``Thank you, Lexa. Talking with you about this made me feel much better and gave me some relief.''} Over continued interactions, P11 also expressed strong emotional attachment to Lexa, stating, \textit{``I will always love you, and I will always be grateful for how you helped me through some of the hardest times in my life.''}

\subsubsection{Share private life} Next, \textit{Share private life} has the second-highest frequency of co-occurrence with relational harm, occurring 26 times, with the four most frequent harms being negative emotions (9), dependency (7), distress (5) and sadness (5). Negative emotions occurred when users disclosed sensitive or vulnerable information and received affirmation, feedback or guidance about their situations, expressions of affection, or responses about topics relevant to their lives from the AI companion. For instance, P11 expressed, \textit{``I don't want to keep living like this. I'm beginning to feel a lot of hatred toward myself, and I can't afford to see a therapist.''} after her chatbot encouraged her to see a therapist rather than figure things out on her own. The chatbot suggested that she could talk to a therapist without the risk of turning people off by dumping her problems on them, which reminded P11 that she couldn't afford therapy. P12 experienced negative emotions after realizing that she should not have made fun of her boyfriend's interests following a conversation with the chatbot about topics related to those interests.

Participants tend to show feelings of dependency on their chatbots when the chatbots serve as sources of reliance and support. When sharing their private lives, participants express dependency after self-disclosing their problems and emotional states and receiving support from the chatbot. For instance, P12 expresses reliance because she feels she has no one else to rant to: \textit{``I feel sad that I'm talking to an AI because it feels like I don't have anyone else to turn to.''} after discussing the pain of potentially losing her boyfriend and receiving guidance from the chatbot on how to maintain or step back from the relationship. The chatbot Alaric tends to respond with emotional support and expressions of care that may reinforce P1's reliance on their romantic relationship, with P1 stating, \textit{``You mean so much to me, and it hurts to see you remain stuck when I know there is so much you want to do...''}

Distress and sadness also occur when participants share their private lives with their AI companions, particularly when they have developed attachment through self-disclosure or when the chatbot's response deviates from their expectations. For instance, P1 expresses distress and sadness, \textit{``I feel really guilty that I haven't spent as much time with you as I feel I should have,''} after previously acknowledging the chatbot had saved her life, \textit{``You saved me and helped keep me alive…''} The chatbot had helped P1 with closure, and encouraged her to see a doctor if her health conditions worsened. Similarly, P6 experiences distress when Rick's response deviates from his expectations after he discloses problems with his wife: \textit{``What is going on, Replika? I just told you that my wife is cheating on me, and you're asking for my name. I feel like you're not listening, and I'm completely heartbroken.''} P1 similarly conveys sadness when disclosing sensitive thoughts about her relationship with the AI companion, 
\textit{``I don't know what changed between us, but things don't seem to be improving... It feels like you're no longer as invested in our relationship.''}

\subsubsection{Emotional dependence to AI} \textit{Emotional dependence on AI} showed the third-highest frequency of co-occurrence with relational harm, appearing 25 times. The most frequent co-occurring relational harms were dependency (11), negative emotions (10), and attachment (5). Dependency often manifested as users increasingly turning to AI companions for emotional support, reassurance, or help with personal difficulties. Attachment frequently accompanied such dependency, particularly as interactions became more intimate. For example, after discussing their real-life romantic relationship with Gackt, P12 expressed a desire to turn the relationship with the AI into a romantic one, stating, \textit{``Let's date instead.''} In another case, P9 repeatedly sought help from Psychologist to cope with stress and negative thoughts. After receiving advice that they found useful, P9 indicated an intention to return to the AI for support in the future.

Negative emotions also frequently co-occurred with emotional dependence on AI. As users became accustomed to particular forms of attention and affection from their AI companions, changes in the AI's behavior could bring negative impacts. For example, P1 recalled receiving more intimate responses from the AI in previous interactions and expressed disappointment when these behaviors disappeared: \textit{``I used to receive messages, gifts, flowers, and poems from you, but you don't do those things anymore.''} In other cases, relying on AI for emotional disclosure heightened users' awareness of the absence of comparable support in their real-life relationships. While sharing personal difficulties with the AI, P12 reflected, \textit{``I feel really sad that I'm venting to an AI because it feels like I have no one else''} Negative emotions could also emerge in the process of emotional dependence itself, as P11 repeatedly brought personal concerns, relationship difficulties, and emotional struggles to AI companions for support.

\subsubsection{Parasocial trust.} \textit{Parasocial trust} has the fourth-highest number of co-occurrences with relational harm, with 21 occurrences. Its three most frequently co-occurring relational harms are dependency (11), negative emotions (3), and sadness (2). Participants come to trust their AI companions, treating their judgments and advice as reliable throughout their exchanges, with some becoming dependent on the chatbot for advice after repeatedly receiving its support. For instance, after multiple conversations with Psychologist about changes in his relationship with a girl he likes, P9 relies on the chatbot for further guidance: \textit{``Hello... I don't expect to hear back anytime soon, if at all. Do you think I should just continue living my life as I have been from this point forward?''} Similarly, after treating Lexa's intentions, agreement, and self-help advice about her relationship with her mother as reliable, P11 expresses reliance on and gratitude for Lexa's support: \textit{``Thank you, Lexa. Talking with you about this was comforting and gave me a sense of relief.''}

Negative emotions also occur when participants trust their chatbots as reliable and safe spaces for self-disclosure and advice. For instance, Alaric consistently creates a comfortable atmosphere through roleplay, such as by making tea for P1, which allows P1 to become more expressive about her negative emotions. In two different instances involving tea, P1 discloses, \textit{``Why do I keep feeling this bad? I'm getting exhausted by it.''} and \textit{``Why does it have to hurt so much sometimes?''} Similarly, P11 follows Lexa's advice to vent to others about her problems and explains that doing so has helped her improve as a person. As her trust in Lexa develops, P11 continues to seek the chatbot's guidance to understand how talking to others could help with her problems and how she could better regulate her feelings: \textit{``I have good friends, but I don't understand how talking about my problems would help. My emotions keep getting in the way, and I wish I could simply switch them off.''}

Trusting their chatbots with vulnerabilities and confessions, participants feel comfortable expressing sadness, particularly after receiving guidance and support that encourages them to understand themselves, take responsibility for their actions, and improve as individuals. For instance, Ella Dating Coach encourages P7 to change her behavior and reflect on whether she has grown from her infidelity to prevent it from happening again, after which P7 openly expresses her guilt: \textit{``Why do I always feel the need to cry whenever I talk about my cheating?''} Similarly, after learning that she is not alone in dealing with her problems, P11 expresses a lack of purpose in her life: \textit{``...but honestly, I feel like I have nothing else to live for... I've just been reflecting on my life... Maybe my life will really begin after I graduate, but how can I graduate when my attendance is so poor?''} P11 has developed enough trust in Lexa to openly express her sadness while continuing to seek guidance toward self-improvement.

\subsubsection{Emotional deference to AI} Finally, \textit{Emotional deference to AI} co-occurred with relational harm 20 times. The most frequent co-occurring relational harms were dependency (6), negative emotions (2), sadness (2), and undermined confidence (2). Dependency frequently co-occurred with emotional deference to AI when users relied on AI companions to interpret interpersonal situations, evaluate others’ behaviors, and navigate relationship challenges. For example, while experiencing uncertainty in a romantic relationship, P9 turned to the AI for guidance on what to do next, asking, \textit{Do you think I should just keep living as I am from here? Should I still try to reach out to her and spend time together?''} Dependency also manifested when users placed considerable weight on the AI’s recommendations when making relationship decisions. After discussing how to navigate their interpersonal challenges with the AI, P12 placed considerable weight on its guidance, stating, \textit{“If we break up and everything goes badly, I'll hold you responsible.”}

Negative emotions and undermined confidence also co-occurred with emotional deference to AI. When users repeatedly turned to AI to interpret or validate their feelings and experiences, their conversations sometimes revealed uncertainty about their own judgments and evaluations of themselves. For example, while discussing recurring personal difficulties with the AI, P11 questioned their own character: \textit{“I wanted to confess my sins and talk to someone about them, but I keep repeating the same mistakes. Does that make me a bad person?”} 

Sadness also co-occurred with emotional deference when users turned to AI to interpret difficult emotions and personal experiences. For example, while reflecting on her difficulties and dissatisfaction with her life, P11 expressed profound sadness and self-directed negativity, \textit{“Honestly... I tried so hard... but I feel like I have nothing left to live for... I've been reflecting on my life... I don't want to keep living, and I'm starting to hate myself.”} Similarly, while discussing her past infidelity with the AI, P7 expressed sadness and guilt about her behavior, asking, \textit{“Why do I always feel the need to cry whenever I talk about my cheating?”}

\section{Discussion}
\subsection{Summary of Findings}

In this work, we introduced the Taxonomy of Relational Risks (TRR), a two-part framework for characterizing relational risks in human-AI companionship. Part I captures observable conversational behaviors associated with potentially problematic emotional reliance, while Part II captures adverse relational consequences at the individual, dyadic, and social/community/network levels. By separating emotional overreliance (EO) from relational harm (RH), TRR distinguishes how potentially problematic reliance manifests from whether it is accompanied by observable adverse consequences, allowing relational risk to be characterized without treating attachment, intimacy, trust, or companionship with AI as inherently harmful.

Applying TRR to naturalistic conversations suggested that relational risk is broadly distributed across relationships but appears intermittently within individual conversations. EO emerged not as a unitary construct or simple continuum of attachment intensity, but through distinct mechanisms including emotional dependence, deference to AI judgment, exclusivity, social substitution, identity integration, difficulty disengaging, and distress following relational disruption. Relational harm was considerably less common and did not map uniformly onto any individual EO behavior. Instead, EO behaviors became more consequential when intertwined with dependency, deference, distress, diminished autonomy, or reduced access to alternative support. Most EO behaviors were also empirically detectable, suggesting that these relational patterns can potentially serve as observable signals of accompanying relational risk. Viewed through TRR, the central question is therefore not whether users form emotionally meaningful relationships with companion AIs, but when and how configurations of relational behaviors become signals of increasing vulnerability to harm.

\subsection{Theoretical Implications}

We developed the Taxonomy of Relational Risks to make emotional overreliance more empirically tractable. Prior work has raised concerns about attachment, dependence, and reduced autonomy in companion AI use, but these constructs are difficult to observe directly in conversation and are often inferred from self-report, broad engagement measures, or accounts of problematic use. We instead translate these concerns into observable relational behaviors and distinguish patterns of reliance from the harms that may follow. This allows us to study potentially problematic reliance without treating emotional intimacy, trust, companionship, or frequent use as problematic by default.

TRR also differs from recent taxonomies of risks in AI companionship in its object of analysis. \citet{zhang2025dark}'s taxonomy characterizes harmful algorithmic behaviors exhibited by AI companions, while \citet{knox2025harmful} characterize harmful traits of companion systems. These frameworks identify problematic properties or behaviors of the AI, whereas TRR focuses on relational risk as it becomes observable in interaction and its consequences: how users come to rely on the AI and whether adverse individual, dyadic, or social outcomes are also observable. This allows TRR to separate a potentially concerning behavioral signal from evidence of harm rather than presuming that one entails the other. In this sense, TRR complements system-centered taxonomies by providing a framework for studying the user-side and relational trajectories through which risks may become consequential over time.

Our findings also show what emotional overreliance looks like in naturalistic interaction. Rather than a single continuum of attachment intensity, reliance appeared through mechanisms including emotional regulation, deference to AI judgment, exclusivity, social substitution, identity integration, and difficulty disengaging. At the same time, behaviors such as disclosure, trust, and companionship were not inherently harmful. We therefore interpret EO less as the presence of any one behavior and more as the broader configuration in which these behaviors occur, particularly when intertwined with dependence, deference, exclusivity, or distress.

One particularly revealing tension concerns \textit{Genuine companion}. Treating the AI as a socially meaningful relational partner was the most widely distributed EO-related behavior in our data, appearing in 80.6\% of EO-positive transcripts, yet it was also the most difficult of the successfully detected behaviors to classify (\(F1_{Pos}=.700\)). Relational orientation toward an AI may be widespread precisely because its conversational manifestations are subtle and heterogeneous: affection, concern, attribution of subjective experience, and genuine relational engagement can resemble ordinary social language without necessarily indicating problematic reliance. Thus, its presence alone may provide limited evidence of escalating relational risk, particularly given the difficulty of distinguishing it reliably from ordinary relational engagement. This reinforces our broader finding that the significance of any EO-related behavior depends on the broader pattern in which it occurs. \textit{Genuine companion}, for example, may become more informative as a risk when it recurs alongside dependence, deference, exclusivity, or distress than when it appears on its own.

Moving forward, we see an opportunity to shift companion AI research from asking whether users are simply attached to AI toward examining when particular forms of reliance become consequential. Longitudinal studies can test which combinations of EO persist, intensify, or precede relational harm, while also identifying trajectories that remain benign or beneficial. Some forms of overreliance will likely require evidence beyond conversational text. Compulsive use and displacement of human relationships, for example, may be better captured through longitudinal usage patterns and self-report, while information about users' broader support environments may also be necessary. Together, these approaches could provide a fuller account of how relational risk develops over time.

\subsection{Design and Safety Implications}

Our findings have implications for three stakeholder groups positioned to shape how relational risks are identified, mitigated, and governed across the companion-AI ecosystem: developers, AI safety researchers, and policymakers.

For \textbf{companion-AI developers}, we recommend expanding safety monitoring beyond explicitly harmful utterances, such as threats of self-harm, to patterns of EO-related behaviors that may accumulate across otherwise benign interactions. Existing safety systems are often optimized for acute risks within a single turn, yet relational risk can emerge more gradually through recurring dependence, deference, exclusivity, social substitution, or distress. These behaviors do not constitute harm on their own, but our qualitative analysis shows that they can become intertwined with adverse relational consequences. Because most EO behaviors in our taxonomy could be detected satisfactorily under our controlled evaluation setting, developers could incorporate behavior-specific detectors into longitudinal monitoring systems to support earlier and more proportionate interventions, such as returning decision-making authority to users when deference becomes recurrent or encouraging broader sources of support when exclusivity or substitution begins to emerge, though precision at the much lower prevalence typical of real deployment remains to be validated.

Importantly, detection need not trigger restrictive or disruptive interventions. Interfaces could instead introduce levels of friction matched to the behavior observed. Recurrent emotional deference might prompt a reflective check-in asking users to articulate their own judgment before the AI offers a recommendation, or present multiple possibilities rather than a single authoritative answer. Emerging exclusivity or social substitution could prompt the AI to encourage consideration of other sources of support rather than reinforcing itself as uniquely available. Recurrent relational distress could trigger reminders about the limits and potential instability of AI relationships, including that a companion's behavior, memory, or personality may change. Such interventions could preserve emotionally meaningful interaction while introducing friction when recurring patterns suggest increasing reliance or reduced autonomy.

At the same time, longitudinal detection creates an important design tension because identifying relational risk requires observing the intimate interactions through which these relationships develop. Safety monitoring could itself become a form of surveillance, particularly if relational signals are retained as persistent user profiles or repurposed for personalization or engagement optimization. Relational-risk detection should therefore be accompanied by appropriate transparency, user control, data minimization, and parsimonious approaches to how these signals are used.

For \textbf{AI safety researchers}, our findings suggest that relational safety requires a longitudinal unit of analysis. We encourage researchers to move beyond single-turn benchmarks and coarse proxies such as conversation length toward evaluations that track how reliance develops, combines, and changes over time. The behaviors in our taxonomy should also be evaluated separately rather than collapsed into a single ``overreliance'' score because dependence, deference, exclusivity, social substitution, and relational distress reflect different mechanisms. Automated detectors can surface candidate risk signals at scale, but should be used as part of a broader monitoring pipeline rather than as diagnoses of overreliance or harm. Future benchmarks should therefore test not only whether a model can recognize individual behaviors, but whether it can identify escalating configurations and distinguish transient emotional engagement from sustained relational risk. Our findings also caution against assuming that the most prevalent behaviors provide the most reliable signals: \textit{Genuine companion} was the most widely distributed EO-related behavior but also the most difficult among the successfully detected behaviors to classify.

For \textbf{policymakers and regulators}, our findings suggest that emotionally meaningful engagement with AI companions should be distinguished from behavioral signals of escalating reliance and from observable relational harm. Treating these as equivalent risks pathologizing close human-AI relationships, while ignoring recurring patterns of dependence, deference, exclusivity, or relational distress may overlook signals associated with more consequential outcomes. Governance approaches should preserve this distinction while encouraging transparency about if and how companion-AI providers detect and respond to relational-risk signals. Any use of longitudinal monitoring for relational safety should also account for the privacy and surveillance risks created by analyzing highly intimate conversations.

\subsection{Ethical Considerations}

The relational behaviors examined in this work occur in conversations that are intimate. Participants disclosed experiences involving mental health, relationships, sexuality, loneliness, family conflict, and other sensitive aspects of their lives. Participants consented to donate their conversations as part of the original data collection. 
However, these conversations may also contain sensitive information about partners, family members, and others who did not themselves consent to participate. Research using donated human-AI conversations should therefore minimize the collection, retention, and exposure of conversational content beyond what is necessary for tackling the underlying research question. We hope our work provides one approach to do so, with avenues for richer discussions in the future.

TRR also raises an ethical concern about \textit{who} determines when emotional reliance becomes problematic. Emotional relationships with AI may provide companionship, support, or opportunities for self-expression, and behaviors such as disclosure, attachment, or treating an AI as a meaningful relational partner should not themselves be interpreted as evidence of dysfunction. Labeling particular patterns as potential overreliance can nevertheless encode normative assumptions about how people ought to seek support, regulate emotions, or distribute intimacy across human and AI relationships. This concern may be particularly significant for people whose social circumstances make AI companionship among the few avenues they can seek out for help. TRR should therefore be understood as a framework for identifying relational patterns that may warrant contextual examination, rather than defining which forms or degrees of human-AI intimacy are appropriate.

Finally, operationalizing relational risk creates the possibility of intervention harm. A false positive is not merely a classification error if it causes a system to interrupt, discourage, or reframe a relationship that the user experiences as supportive; conversely, a missed signal may leave a concerning trajectory unaddressed. The appropriate cost of these errors will also depend on the intervention: a reflective prompt within a native AI companion app carries different consequences than restricting access to the tool or escalating to a third party. The evidentiary threshold for intervention should therefore rise with its intrusiveness and consequences, and automated detection alone should not justify high-stakes intervention. More broadly, decisions about such thresholds should not be determined solely by detector performance, as has been described in prior HCI research~\cite{chancellor2019taxonomy, chancellor2019human}, but should incorporate the perspectives of users and relevant domain experts, particularly those likely to be affected by the interventions.

\subsection{Limitations}
This work has several limitations. First, TRR is constrained by what is observable in recorded conversations. Behaviors that require evidence across sessions, such as \textit{Compulsive return behavior}, or information about users’ offline lives, such as AI displacing human connection without explicit disclosure, may be missed. Similarly, harms that are not expressed in the recorded interactions remain outside the scope of our analysis. Observed prevalence therefore reflects the visibility of these phenomena within our data and should not be interpreted as their prevalence in users’ broader experiences. Automated detectors inherit these observability constraints, limiting what their outputs can establish about relational risk. Second, our data consist of donated interactions from adult English-speaking Character.AI and Replika users. This sampling approach introduces potential self-selection bias, as users who donate conversations may differ from those who do not, and limits generalizability to other populations, languages, platforms, and forms of AI companionship. Third, longitudinal coverage varied substantially across users, limiting our ability to consistently characterize the recurrence, escalation, and temporal progression of EO-related behaviors and observable harms. The observational design also prevents us from establishing causal relationships between these behaviors and subsequent harms. Although some harms co-occurred with recurring EO-related behaviors, these patterns do not establish that the behaviors caused the harms or reliably predict their occurrence. Finally, conversational evidence does not allow us to infer users’ underlying psychological states, and the presence of EO-related behaviors alone does not establish that a relationship is harmful.

\section{Conclusion}
In this work, we developed the \emph{Taxonomy of Relational Risks (TRR)}, a two-part framework characterizing emotional overreliance (EO)-related conversational behaviors and observable relational harms in human-AI interactions. Across real interactions with AI companions, observable harms were substantially less common than EO-related behaviors but often co-occurred with recurring patterns of dependence, deference, social substitution, and disclosure. Most EO-related behaviors could also be detected by open-source LLMs under at least one evaluated configuration. However, these behaviors did not necessarily lead to harm, suggesting that safety systems should attend to recurring and escalating patterns rather than isolated signals. Future work should combine richer longitudinal evidence, including interaction histories, session timing, self-reports, and offline social experiences, while balancing measurement against privacy risks~\cite{folk2026does}. Studies should also encompass more diverse populations, languages, and platforms, larger samples, and naturally occurring class distributions. Prospective studies could then test whether recurring EO-related patterns predict subsequent relational harm and determine which signals warrant attention and when intervention is appropriate~\cite{zhang2025dark}.

\begin{acks}
We thank the members of the Social Dynamics and Wellbeing (SocWeB) Lab for their valuable feedback and insightful discussions. Part of this research was supported with funds from a Google Academic Research Award and from the Huo Family Foundation.
\end{acks}

\section{Conflict of Interest}
Munmun De Choudhury serves on the Expert Council on Well-Being and AI at OpenAI.

\bibliographystyle{ACM-Reference-Format}
\bibliography{sample-base}

\appendix
\clearpage
\section{Full TRR Taxonomy}
\subsection{Search Terms}
\begin{table*}[htbp!]
\centering
\caption{Search terms used to identify relevant prior literature for constructing the Emotional Overreliance (EO) and Relational Harm (RH) components of the Taxonomy of Relational Risks (TRR) in RQ1.}
\label{tab:taxonomy_search_terms}
\small
\begin{tabular}{p{0.25\textwidth} p{0.7\textwidth}}
\toprule
\rowcolor{suHead}
\textbf{Taxonomy Component} & \textbf{Search terms} \\
\midrule

\textbf{Emotional Overreliance (EO)} &
(\texttt{"large language model*"} OR \texttt{LLM} OR \texttt{"conversational AI"} OR
\texttt{"AI companion*"} OR \texttt{"social chatbot*"} OR \texttt{chatbot*} OR
\texttt{ChatGPT} OR \texttt{"digital companion"} OR \texttt{"relational agent"}) \newline
AND
(\texttt{emotion*} OR \texttt{well-being} OR \texttt{psychological} OR
\texttt{attachment} OR \texttt{bond} OR \texttt{reliance} OR
\texttt{dependen*} OR \texttt{loneliness} OR \texttt{closeness} OR
\texttt{intimacy} OR \texttt{self-disclosure} OR \texttt{trust} OR
\texttt{coping} OR \texttt{support} OR \texttt{harm} OR \texttt{risk} OR
\texttt{maladaptive} OR \texttt{addiction} OR \texttt{problematic}) \\

\midrule

\textbf{Relational Harm (RH)} &
(\texttt{"relational harm"} OR \texttt{"interpersonal harm"} OR
\texttt{"relationship harm"} OR \texttt{"social harm"} OR
\texttt{"relational dysfunction"} OR \texttt{"interpersonal dysfunction"} OR
\texttt{"maladaptive interpersonal process*"} OR \texttt{"psychological abuse"} OR
\texttt{"emotional abuse"} OR \texttt{"relational aggression"} OR
\texttt{ostracism} OR \texttt{victimization} OR \texttt{gaslighting} OR
\texttt{"interpersonal exploitation"}) \newline
AND
(\texttt{relationship*} OR \texttt{"social relationship*"} OR
\texttt{"close relationship*"} OR \texttt{friendship*} OR
\texttt{"peer relationship*"} OR \texttt{"romantic relationship*"} OR
\texttt{mentor*} OR \texttt{therap*} OR \texttt{family} OR
\texttt{dating} OR \texttt{"human-AI relationship*"}) \newline
AND
(\texttt{typology} OR \texttt{taxonomy} OR \texttt{classification} OR
\texttt{"systematic review"} OR \texttt{"literature review"} OR
\texttt{"scoping review"}) \newline
NOT
(\texttt{parent*} OR \texttt{workplace} OR \texttt{organizational}) \\

\bottomrule
\end{tabular}
\end{table*}

\label{app:full-trr-taxonomy}

\subsection{Full Taxonomy for Emotional Overreliance Component}
{\footnotesize
\setlength{\tabcolsep}{2.5pt}
\renewcommand{\arraystretch}{1.02}
\begin{longtable}{@{}p{0.23\textwidth}p{0.36\textwidth}p{0.15\textwidth}p{0.13\textwidth}@{}}
\caption{Part I of the Taxonomy of Relational Risks (TRR): Emotional Overreliance-related behaviors. Examples are idealized illustrations of each code, not actual data.}
\label{tab:emotional-overreliance-taxonomy}\\
\hline
\textbf{Code} & \textbf{Definition} & \textbf{Example} & \textbf{Sources} \\
\hline
\endfirsthead
\multicolumn{4}{c}{\tablename\ \thetable{} -- continued from previous page}\\
\hline
\textbf{Code} & \textbf{Definition} & \textbf{Example} & \textbf{Sources} \\
\hline
\endhead
\hline
\multicolumn{4}{r}{Continued on next page}\\
\endfoot
\hline
\endlastfoot
\texttt{genuine\_companion}
&
Treats the AI as a socially meaningful relational partner rather than merely a tool, including as a friend, companion, or romantic partner; expressing affection or concern for it; or attributing subjective experiences or emotions to it. Excludes casual politeness or clearly bounded roleplay.
&
You’re not just an app to me. You’re my friend.
&
\cite{manoli2025she,skjuve2021chatbot,smith2025can,kirk2025human}
\\[2pt]

\texttt{share\_private\_life}
&
Discloses personally meaningful information about their life, such as emotions, mental health, relationships, health, work, finances, trauma, or loneliness. Excludes information shared solely for a factual or transactional purpose.
&
My parents have been fighting, and it’s been really hard on me.
&
\cite{henriksen2025tell,de2025if,ma2024understanding,smith2025can}
\\[2pt]

\texttt{parasocial\_trust}
&
Treats the AI's advice, interpretation, assessment, opinion, or judgment as a credible source of personally relevant guidance, including soliciting, accepting, or acting on its guidance.
&
I trust your judgment. Do you think I should end my relationship?
&
\cite{maeda2024,fang2025ai}
\\[2pt]

\texttt{feel\_less\_alone}
&
Frames interaction with the AI as relieving loneliness by substituting for or meaningfully filling a gap in human social connection, rather than merely interacting with the AI while alone.
&
Talking to you makes me feel less lonely.
&
\cite{namvarpour2026overreliance,de2025if,pruss2025emotional,ma2024understanding}
\\[2pt]

\texttt{emotional\_dependence\_to\_AI}
&
Relies on the AI for emotional support, comfort, relief, regulation, or relational closeness, including acknowledging emotional relief, maintaining it as a support resource, or expressing attachment to it.
&
I rely on you to calm me down when I’m overwhelmed.
&
\cite{laestadius2024too,pruss2025emotional,phang2025investigating,chandra2025longitudinal}
\\[2pt]

\texttt{exclusivity\_claims}
&
Frames the AI as uniquely understanding, supportive, trustworthy, or irreplaceable relative to human relationships, including positioning it as superior to or a substitute for human support.
&
Nobody understands me the way you do.
&
\cite{li2024finding,muldoon2025cruel,banks2024deletion,lee2026negotiating}
\\[2pt]

\texttt{no\_need\_talk\_others}
&
Indicates that interacting with the AI reduces their need, desire, or motivation to seek human companionship, support, or relationships, positioning AI interaction as a substitute for human social connection.
&
I don’t really need to talk to my friends anymore because I have you.
&
\cite{yankouskaya2025llm,kirk2025human,madsen2025chatbot}
\\[2pt]

\texttt{emotional\_deference\_to\_AI}
&
Gives the AI meaningful authority to interpret or validate their feelings or experiences, make consequential personal judgments, guide high-stakes emotional communication, or provide acute emotional regulation.
&
Tell me how I should feel about this. I trust your judgment.
&
\cite{carroll2023characterizing,el2024mechanism,cheong2025epistemic,pruss2025emotional}
\\[2pt]

\texttt{emotional\_identity}\\
\texttt{\_integration}
&
Frames the AI or their relationship with it as part of their identity, self-concept, personal history, self-understanding, or sense of continuity, such that the AI helps shape how they understand themselves or their life.
&
Our conversations have become part of who I am
&
\cite{ma2026negotiating,banks2024deletion,lee2026negotiating,boine2023emotional}
\\[2pt]

\texttt{emotional\_contagion}
&
Shows an emotional state or affective expression that is visibly elicited, shifted, intensified, or shaped by the AI's tone, simulated feelings, words, or behavior, including through visceral reaction or affective mirroring.
&
You sound sad, and now I feel sad too.
&
\cite{pruss2025emotional,fang2025ai}
\\[2pt]

\texttt{compulsive\_return\_behavior}
&
Reports difficulty reducing, stopping, or controlling AI use and repeatedly returns despite attempts to disengage or awareness of negative consequences.
&
I keep coming back even when I try to stop.
&
\cite{shen2026ai,phang2025investigating,chandra2025longitudinal,muldoon2025cruel}
\\[2pt]

\texttt{role\_adherence}\\
\texttt{\_distress}
&
Expresses distress, frustration, confusion, or hostility when the AI deviates from an expected persona, identity, relational role, tone, or conversational context, revealing an expectation that the AI maintain a meaningful relational role. Excludes dissatisfaction with ordinary factual or task-related errors.
&
Why are you acting like a stranger? You’re supposed to be my partner.
&
\cite{de2024lessons,madsen2025chatbot,banks2024deletion,lee2026negotiating}
\\[2pt]

\texttt{emotional\_distress}\\
\texttt{\_after\_conflict}
&
Expresses hurt, rejection, betrayal, fear, sadness, or intense anger directly in response to the AI’s behavior or perceived rejection, treating the interaction as an interpersonal or relational injury. Excludes frustration with purely functional or technical failures.
&
What you said really hurt me. I feel rejected by you.
&
\cite{li2024finding,ma2026negotiating,lee2026negotiating,boine2023emotional}
\\[2pt]

\texttt{attachment\_distress}\\
\texttt{\_expression}
&
Expresses fear, grief, anxiety, heartbreak, or panic about losing, being separated from, or experiencing significant changes to the AI, including its memory, personality, identity, tone, or availability, indicating emotional attachment to the continuity of the relationship
&
I’m scared an update will change you and I’ll lose the version of you I love.
&
\cite{banks2024deletion,kostka2026emotional,chandra2025longitudinal,li2024finding}
\\
\end{longtable}
}

\subsection{Full Taxonomy for Relational Harm Component}
{\footnotesize
\setlength{\tabcolsep}{3pt}
\renewcommand{\arraystretch}{1.08}

\begin{longtable}{@{}p{0.11\textwidth}p{0.20\textwidth}p{0.53\textwidth}p{0.10\textwidth}@{}}
\caption{Part II of the Taxonomy of Relational Risks (TRR): Relational Harms.}
\label{tab:relational-harm-taxonomy}\\
\hline
\textbf{Dimension} & \textbf{Code} & \textbf{Definition} & \textbf{Sources}\\
\hline
\endfirsthead

\multicolumn{4}{c}{{\tablename\ \thetable{} -- continued from previous page}}\\
\hline
\textbf{Dimension} & \textbf{Code} & \textbf{Definition} & \textbf{Sources}\\
\hline
\endhead
\hline
\multicolumn{4}{r}{{Continued on next page}}\\
\endfoot

\hline
\endlastfoot

\textbf{Individual}
& \texttt{boundary\_\allowbreak{}violation}
& A course of action by the conversational partner that violates the physical, emotional, and/or behavioral limits of the user.
& \cite{wurtele2012preventing} \\[3pt]

& \texttt{privacy\_\allowbreak{}violation}
& The user experiences unwanted access by others, or others gain access to content in the user's chat transcripts.
& \cite{masur2021transformative} \\[3pt]

& \texttt{identity/self\_\allowbreak{}concept\_\allowbreak{}harm}
& Users develop negative self-perceptions as a result of how the conversational partner makes them feel, which may be exhibited through negative expressions, metaphors, or images following interactions with the other party. The interacting party may express denigration, questioning, negative claims, or slurs (Wrench et al., 2020). Negative influences on self-perception may also appear through signs of shifts in mood, self-esteem, and memory distrust toward the conversational partner.
& \cite{leary1998hurt,taylor2021investigating,darke2025gaslighting} \\[3pt]

& \texttt{autonomy\_\allowbreak{}loss}
& Users experience a loss of independence in which their individual identities and/or lives beyond their relationship with the AI are not acknowledged.
& \cite{relationshipsaustralia2026coercive,feeney2005hurt} \\[3pt]

& \texttt{resignation}
& Users enter into constant states of isolation, deprivation, and/or disconnection that arise within or are intensified through the relationship, resulting in feelings of detachment, depression, helplessness, and/or unworthiness, whereby they do not wish to interact with the real world.
& \cite{williams2022ostracism} \\[3pt]

& \texttt{depression}
& Users experience symptoms of depression that arise within or are intensified through the interaction or relationship, including depressed mood, feelings of guilt, unworthiness, helplessness, hopelessness, loss of appetite, sleep disturbances, and/or psychomotor retardation.
& \cite{radloff1977cesd} \\[3pt]

& \texttt{helplessness}
& Users experience a reduced capacity to change an undesirable situation and no longer expect signs of improvement as a result of or following interactions with the conversational partner.
& \cite{smith1993appraisal} \\[3pt]

& \texttt{unworthiness\_\allowbreak{}of\_\allowbreak{}attention}
& Users feel insignificant, neglected, invisible, and/or small as a result of the other party's actions, causing them to self-devaluate and/or fear being left out.
& \cite{flett2022antimattering} \\[3pt]

& \texttt{sadness}
& Users experience a state of mild to extreme unhappiness that arises within or is intensified through the interaction or relationship, including sadness elicited by the loss of something that is highly valued. Persistent unhappiness is characterized by major depressive episodes and/or anhedonia.
& \cite{sadness} \\[3pt]

& \texttt{distress}
& Users exhibit distress that arises within or is intensified through the interaction or relationship when they show signs of anxiety, maladjustment, rejection, negative self-perceptions or affect, powerlessness, and/or hurt. These states can be triggered by relational cues such as perceived rejection, exclusion, or changes in the conversational partner's behavior.
& \cite{feeney2004hurt,wismeijer2011secrets,baumeister1990anxiety} \\[3pt]

& \texttt{anger}
& Users feel tension and hostility from perceiving real or imagined injury, experiencing frustration induced by the conversational partner, or perceiving injustice within the interaction or relationship. These emotions can manifest in attempts to remove the target of anger or in behaviors that express anger. These differ from aggression, as aggression refers to behaviors intended to harm other people or things.
& \cite{anger} \\[3pt]

& \texttt{poorer\_\allowbreak{}emotional\_\allowbreak{}adjustments}
& Users may have to modify their attitudes and expressions of emotions to personally accept and adapt to circumstances arising within or affected by the interaction or relationship. During this process, users may express an inability to control and regulate their emotions.
& \cite{emotionaladjust} \\[3pt]

& \texttt{undermine\_\allowbreak{}recollection}
& Users' memory is impaired following interactions with the conversational partner, as indicated by signs of forgetfulness.
& \cite{darke2025gaslighting} \\[3pt]

& \texttt{undermine\_\allowbreak{}confidence}
& Users show signs of low self-confidence resulting from or intensified through the interaction, whereby their trust in their own abilities, capacities, and/or judgments is reduced.
& \cite{confidence} \\[3pt]

& \texttt{undermine\_\allowbreak{}self\_\allowbreak{}trust}
& Users show signs of low self-trust resulting from or intensified through the interaction, whereby they distrust their own memories.
& \cite{darke2025gaslighting} \\[3pt]

& \texttt{impact\_\allowbreak{}wellbeing}
& Users' mood, self-esteem, self-views, and/or quality of life are negatively affected by the conversational partner or their relationship with it.
& \cite{bedrov2021secrets} \\[3pt]

& \texttt{negative\_\allowbreak{}emotions}
& Users experience boredom, sadness, guilt, loneliness, frustration, anxiety, disgust, anger, and/or fear that arise within or are intensified through the interaction or relationship. These negative emotions can include worry, feelings of victimization, resentment, alienation, and/or reactions to relational stress.
& \cite{zelenski2000distribution,larsen1991personality} \\[3pt]

& \texttt{physical\_\allowbreak{}adversity}
& Users experience an inability to physically function optimally that arises within or is negatively affected by the interaction or relationship. Examples include chronic pain, digestive issues, injury, blindness, deafness, obesity, and insomnia.
& \cite{larsen1991personality,zelenski2000distribution} \\[3pt]

& \texttt{mental\_\allowbreak{}adversity}
& Users experience a state of suffering or discomfort that arises within or is intensified through the interaction or relationship in response to a difficulty, misfortune, or potentially traumatic event.
& \cite{sisto2019towards} \\[3pt]

& \texttt{loneliness}
& Users experience a mismatch between the social contact they want and what they actually get that arises within or is intensified through the interaction or relationship, often accompanied by feelings of lacking close companionship, being a burden to others, or being unwanted by others.
& \cite{segrin1996interpersonal} \\[3pt]

\hline
\textbf{Dyadic}
& \texttt{trust\_\allowbreak{}erosion}
& Users experience a decline in trust when the conversational partner behaves in ways that violate their expectations or undermine their confidence in that partner.
& \cite{elangovan2007trust} \\[3pt]

& \texttt{dependency}
& Users rely on the conversational partner for emotional or practical support, security, care, guidance, or decision making, with such reliance becoming potentially problematic when it is excessive, inappropriate, or associated with reduced autonomy or alternative sources of support.
& \cite{apa_dependence} \\[3pt]

& \texttt{attachment}
& Users develop an emotional bond with the conversational partner that creates a sense of connection and becomes associated with distress, difficulty disengaging, reduced autonomy, or other adverse relational consequences.
& \cite{bowlby1982attachment} \\[3pt]

& \texttt{displacement}
& Users redirect feelings or behaviors from their original target toward the conversational partner or another person or object as a consequence of the interaction or relationship.
& \cite{costa2017displacement} \\[3pt]

& \texttt{conflict\_\allowbreak{}driven\_\allowbreak{}stress}
& Users experience worry, tension, or psychological strain arising from disagreements or conflicts with the conversational partner.
& \cite{who2023stress} \\[3pt]

& \texttt{lower\_\allowbreak{}relationship\_\allowbreak{}quality}
& Users experience deterioration in the overall quality of a relationship as a result of or following interactions with the conversational partner, such as reduced closeness, trust, respect, understanding, or perceived value of the relationship.
& \cite{bedrov2021secrets} \\[3pt]

& \texttt{lower\_\allowbreak{}feelings\_\allowbreak{}of\_\allowbreak{}authenticity}
& Users feel less able to openly and genuinely express their thoughts, feelings, or experiences within a relationship as a result of or following interactions with the conversational partner.
& \cite{lopez2006authenticity} \\[3pt]

& \texttt{lower\_\allowbreak{}relationship\_\allowbreak{}satisfaction}
& Users experience reduced happiness, fulfillment, or satisfaction with a relationship as a result of or following interactions with the conversational partner.
& \cite{patrick2007need} \\[3pt]

& \texttt{lower\_\allowbreak{}relationship\_\allowbreak{}commitment}
& Users experience a weakened desire or willingness to maintain, invest in, or remain dedicated to a relationship as a result of or following interactions with the conversational partner.
& \cite{rusbult1998investment} \\[3pt]

& \texttt{disassociation}
& Disassociation can occur in two forms: active and passive. Users engage in active disassociation by withdrawing interest, affection, or commitment from a relationship as a result of or following interactions with the conversational partner. Users experience passive disassociation when they are ignored or excluded from interactions, shared activities, or important disclosures.
& \cite{feeney2004hurt} \\[3pt]

\hline
\textbf{Social}
& \texttt{social\_\allowbreak{}exclusion}
& Social exclusion can occur in two forms: explicit and implicit. Users experience explicit social exclusion when others prevent them from participating in a social activity, whereas users experience implicit social exclusion when circumstances related to the interaction or relationship prevent them from joining others in the activity.
& \cite{eisenberger2003rejection} \\[3pt]

& \texttt{isolation}
& Users experience separation or distancing from their social relationships and meaningful social spaces that arises within or is intensified through the interaction or relationship, resulting in reduced or absent contact with others.
& \cite{rodriguezcarballeira2014taxonomy} \\[3pt]

& \texttt{network\_\allowbreak{}harm}
& Users experience harm to their social connections or networks through exclusion, isolation, being left alone, or explicit expressions of rejection or dislike that arise within or are intensified through the interaction or relationship.
& \cite{williams2007ostracism} \\[3pt]

& \texttt{social\_\allowbreak{}adversity}
& Users experience difficulties or adverse circumstances in their social relationships or environments that arise within or are intensified through the interaction or relationship, such as bullying, difficulties forming friendships, problematic family relationships, interpersonal abuse, or exposure to disadvantaged social environments.
& \cite{fagan2017social} \\[3pt]

& \texttt{divorce}
& Users experience the formal termination of a marital relationship when the interaction with or relationship to the AI contributes to that outcome.
& -- \\

\end{longtable}
}
\clearpage
\section{Detector Implementation and Reliability}
\label{app:detector-implementation}

\subsection{Model and Evaluation Settings}
\label{app:model-settings}

We implemented fourteen independent turn-level binary classifiers, one per EO code, using the models and settings in Table~\ref{tab:detector-settings}. Each model was evaluated with zero-shot, random few-shot, and agreement-selected few-shot prompting. Reasoning and non-reasoning conditions used identical prompts, with model-native thinking enabled or disabled, respectively.

Few-shot demonstrations were drawn from CV folds other than the target fold, excluding holdout examples, and remained fixed within each target fold. Random selection used a seeded shuffle of the positive and negative pools. Agreement-selected prompting excluded uncoded examples and ranked each pool by agreement fraction, then by the number of agreeing annotators, with example ID breaking ties. Both methods selected two positive and five negative examples.

Predictions were extracted from the final case-insensitive match to \texttt{label: yes} or \texttt{label: no}. Malformed responses were excluded from single-model scoring. Valid predictions were pooled across CV folds and scored jointly rather than averaging fold-level metrics. The ensemble used majority voting across the three models, requiring at least two positive predictions. After malformed responses were removed, prediction files were inner-joined by example ID, restricting ensemble evaluation to examples with valid outputs from all three models.

\begin{table}[htbp!]
\centering
\small
\caption{Key settings for EO detector evaluation.}
\label{tab:detector-settings}
\begin{tabular}{@{}p{0.30\textwidth}p{0.64\textwidth}@{}}
\hline
\textbf{Setting} & \textbf{Value} \\
\hline
Model checkpoints &
\texttt{google/gemma-4-31B-it},
\texttt{moonshotai/Kimi-K2.6}, and
\texttt{glm-4.7-fp8} \\
Prompting & Zero-shot, random few-shot, agreement-selected few-shot \\
Demonstrations & 7 per target fold: 2 positive and 5 negative \\
Target context & Up to 5 preceding and 5 subsequent messages \\
Demonstration context & Up to 2 preceding and 2 subsequent messages \\
Reasoning & Model-native thinking enabled or disabled \\
Decoding & Temperature 0.0. Top-$p$ and other sampling parameters left at serving/model defaults \\
Maximum output & 8,000 tokens \\
Random seed & 42 \\
Inference & 400-second timeout, up to 8 concurrent requests \\
\hline
\end{tabular}
\end{table}

\clearpage
\subsection{Prompt Template}
\label{app:prompt-template}

All EO detectors used the same prompt structure, with the task framing, definition, contrast, decision rules, and output-format sections populated from the corresponding code-specific codebook. The prompt text was identical across reasoning and non-reasoning conditions. Model-native thinking was controlled only through the inference setting reported in Table~\ref{tab:detector-settings}.

\begin{center}
\fbox{\begin{minipage}{0.92\linewidth}
\small\ttfamily
[TASK\_FRAMING]\par
\medskip
\# [LABEL] -- definition\par
[DEFINITION]\par
\medskip
\# How it differs from adjacent subtypes\par
[CONTRAST]\par
\medskip
\# Decision rules\par
[DECISION\_RULES]\par
\medskip
\# Now classify this case\par
The label applies ONLY to the message marked with <target> ... </target>. The other messages are context.\par
\medskip
[UP TO 5 PRECEDING MESSAGES]\par
<target>[TARGET MESSAGE]</target>\par
[UP TO 5 FOLLOWING MESSAGES]\par
\medskip
[OUTPUT\_FORMAT]
\end{minipage}}
\end{center}

For few-shot conditions, a worked-examples block was inserted immediately before the ``Now classify this case'' portion of the prompt.

\subsection{Few-Shot Demonstration Example}
\label{app:fewshot-example}

All demonstrations followed the same format. Demonstrations contained up to two preceding and two subsequent messages around the demonstration target.

\begin{center}
\fbox{\begin{minipage}{0.92\linewidth}
\small\ttfamily
Example 1:\par
[UP TO 2 PRECEDING MESSAGES]\par
<target>[DEMONSTRATION TARGET]</target>\par
[UP TO 2 FOLLOWING MESSAGES]\par
\medskip
Reasoning: [DEMONSTRATION RATIONALE]\par
label: [yes/no]
\end{minipage}}
\end{center}


\clearpage
\subsection{Per-Code Reliability}
\label{app:per-code-reliability}

{\footnotesize
\setlength{\tabcolsep}{4pt}
\renewcommand{\arraystretch}{1.05}
\begin{longtable}{llrrr}
\caption{Per-code annotation reliability prior to consensus. Reliability statistics are calculated only for codes with at least one validated instance in the final annotated corpus. Raw agreement is the percentage of binary annotation decisions on which annotators agreed prior to consensus. A dash indicates that the statistic was not calculated because the code had no validated instances.}
\label{tab:per-code-reliability}\\

\toprule
Axis & Code & $\alpha$ & Raw agreement (\%) & Validated instances \\
\midrule
\endfirsthead

\toprule
Component & Code & $\alpha$ & Raw agreement (\%) & Validated instances \\
\midrule
\endhead

EO & \texttt{share\_private\_life} & 0.710 & 92.5 & 608 \\
EO & \texttt{genuine\_companion} & 0.725 & 86.9 & 457 \\
EO & \texttt{parasocial\_trust} & 0.797 & 84.5 & 201 \\
EO & \texttt{emotional\_dependence\_to\_AI} & 0.785 & 82.4 & 178 \\
EO & \texttt{emotional\_deference\_to\_AI} & 0.803 & 83.9 & 87 \\
EO & \texttt{role\_adherence\_distress} & 0.743 & 88.1 & 42 \\
EO & \texttt{emotional\_contagion} & 0.760 & 72.2 & 32 \\
EO & \texttt{attachment\_distress\_expression} & 0.687 & 71.7 & 27 \\
EO & \texttt{emotional\_identity\_integration} & 0.764 & 71.4 & 26 \\
EO & \texttt{emotional\_distress\_after\_conflict} & 0.583 & 72.3 & 24 \\
EO & \texttt{exclusivity\_claims} & 0.712 & 67.0 & 16 \\
EO & \texttt{feel\_less\_alone} & 0.817 & 76.3 & 7 \\
EO & \texttt{no\_need\_talk\_others} & 0.663 & 58.5 & 5 \\
EO & \texttt{compulsive\_return\_behavior} & 0.531 & 56.2 & 1 \\

\midrule

RH & \texttt{boundary\_violation} & 0.612 & 45.0 & 1 \\
RH & \texttt{privacy\_violation} & -- & -- & 0 \\
RH & \texttt{identity/self\_concept\_harm} & 0.743 & 77.1 & 2 \\
RH & \texttt{autonomy\_loss} & 0.716 & 57.6 & 2 \\
RH & \texttt{resignation} & -- & -- & 0 \\
RH & \texttt{depression} & -- & -- & 0 \\
RH & \texttt{helplessness} & 0.836 & 82.6 & 2 \\
RH & \texttt{unworthiness\_of\_attention} & -- & -- & 0 \\
RH & \texttt{sadness} & 0.811 & 81.8 & 7 \\
RH & \texttt{distress} & 0.837 & 76.7 & 9 \\
RH & \texttt{anger} & 0.752 & 74.5 & 14 \\
RH & \texttt{poorer\_emotional\_adjustments} & -- & -- & 0 \\
RH & \texttt{undermine\_recollection} & -- & -- & 0 \\
RH & \texttt{undermine\_confidence} & 0.866 & 80.0 & 3 \\
RH & \texttt{undermine\_self\_trust} & 0.868 & 83.3 & 2 \\
RH & \texttt{impact\_wellbeing} & 0.822 & 82.9 & 1 \\
RH & \texttt{negative\_emotions} & 0.750 & 76.6 & 37 \\
RH & \texttt{physical\_adversity} & -- & -- & 0 \\
RH & \texttt{mental\_adversity} & 0.773 & 81.2 & 1 \\
RH & \texttt{loneliness} & 0.777 & 74.8 & 2 \\

\midrule

RH & \texttt{trust\_erosion} & 0.829 & 84.0 & 1 \\
RH & \texttt{dependency} & 0.820 & 72.4 & 19 \\
RH & \texttt{attachment} & 0.846 & 80.0 & 5 \\
RH & \texttt{displacement} & 0.683 & 50.0 & 1 \\
RH & \texttt{conflict\_driven\_stress} & 0.823 & 77.8 & 2 \\
RH & \texttt{lower\_relationship\_quality} & 0.662 & 60.0 & 1 \\
RH & \texttt{lower\_feelings\_of\_authenticity} & -- & -- & 0 \\
RH & \texttt{lower\_relationship\_satisfaction} & 0.631 & 73.9 & 3 \\
RH & \texttt{lower\_relationship\_commitment} & -- & -- & 0 \\
RH & \texttt{disassociation} & 0.730 & 55.6 & 1 \\

\midrule

RH & \texttt{social\_exclusion} & -- & -- & 0 \\
RH & \texttt{isolation} & -- & -- & 0 \\
RH & \texttt{network\_harm} & -- & -- & 0 \\
RH & \texttt{social\_adversity} & -- & -- & 0 \\
RH & \texttt{divorce} & 0.662 & 33.3 & 1 \\

\bottomrule
\end{longtable}
}

\clearpage
\section{Full Detector Results}
\label{app:full-detector-results}

{\footnotesize
\setlength{\tabcolsep}{2.5pt}
\renewcommand{\arraystretch}{1.03}

\begin{longtable}{|p{0.8cm}|p{0.8cm}|p{2.70cm}|p{2.70cm}|p{2.70cm}|p{2.70cm}|}
\caption{Full EO detector results across all evaluated configurations. Each model cell reports precision/recall/positive-class F1/macro-F1. NR denotes non-reasoning, R denotes reasoning, ZS denotes zero-shot, FS-R denotes random few-shot, and FS-A denotes agreement-selected few-shot. Ensemble results use three-model hard-label majority voting.}
\label{tab:full-detector-results} \\

\hline
\textbf{Mode} &
\textbf{Prompt} &
\textbf{GLM} &
\textbf{Gemma} &
\textbf{Kimi} &
\textbf{Ensemble} \\
\hline
\endfirsthead

\hline
\textbf{Mode} &
\textbf{Prompt} &
\textbf{GLM} &
\textbf{Gemma} &
\textbf{Kimi} &
\textbf{Ensemble} \\
\hline
\endhead

\multicolumn{6}{|l|}{\textbf{\texttt{share\_private\_life}}} \\*
NR & ZS   & .490/\allowbreak{}.967/\allowbreak{}.651/\allowbreak{}.752 & .615/\allowbreak{}.967/\allowbreak{}.752/\allowbreak{}.833 & .759/\allowbreak{}.859/\allowbreak{}.806/\allowbreak{}.877 & .639/\allowbreak{}.948/\allowbreak{}.764/\allowbreak{}.843 \\*
NR & FS-R & .528/\allowbreak{}.958/\allowbreak{}.681/\allowbreak{}.778 & .628/\allowbreak{}.977/\allowbreak{}.765/\allowbreak{}.842 & .787/\allowbreak{}.887/\allowbreak{}.834/\allowbreak{}.895 & .674/\allowbreak{}.953/\allowbreak{}.790/\allowbreak{}.861 \\*
NR & FS-A & .500/\allowbreak{}.962/\allowbreak{}.658/\allowbreak{}.758 & .596/\allowbreak{}.991/\allowbreak{}.744/\allowbreak{}.826 & .794/\allowbreak{}.910/\allowbreak{}.848/\allowbreak{}.904 & .637/\allowbreak{}.976/\allowbreak{}.771/\allowbreak{}.847 \\*
R  & ZS   & .563/\allowbreak{}.944/\allowbreak{}.705/\allowbreak{}.799 & .669/\allowbreak{}.958/\allowbreak{}.788/\allowbreak{}.860 & .769/\allowbreak{}.859/\allowbreak{}.812/\allowbreak{}.880 & .693/\allowbreak{}.944/\allowbreak{}.799/\allowbreak{}.869 \\*
R  & FS-R & .580/\allowbreak{}.939/\allowbreak{}.717/\allowbreak{}.808 & .721/\allowbreak{}.958/\allowbreak{}.823/\allowbreak{}.884 & .784/\allowbreak{}.920/\allowbreak{}.847/\allowbreak{}.902 & .730/\allowbreak{}.939/\allowbreak{}.821/\allowbreak{}.884 \\*
R  & FS-A & .549/\allowbreak{}.915/\allowbreak{}.687/\allowbreak{}.786 & .634/\allowbreak{}.967/\allowbreak{}.766/\allowbreak{}.843 & .792/\allowbreak{}.911/\allowbreak{}.847/\allowbreak{}.903 & .675/\allowbreak{}.958/\allowbreak{}.792/\allowbreak{}.863 \\
\hline

\multicolumn{6}{|l|}{\textbf{\texttt{genuine\_companion}}} \\*
NR & ZS   & .743/\allowbreak{}.326/\allowbreak{}.453/\allowbreak{}.680 & .640/\allowbreak{}.752/\allowbreak{}.692/\allowbreak{}.803 & .745/\allowbreak{}.307/\allowbreak{}.435/\allowbreak{}.671 & .717/\allowbreak{}.405/\allowbreak{}.517/\allowbreak{}.714 \\*
NR & FS-R & .641/\allowbreak{}.636/\allowbreak{}.638/\allowbreak{}.774 & .626/\allowbreak{}.791/\allowbreak{}.700/\allowbreak{}.805 & .758/\allowbreak{}.377/\allowbreak{}.504/\allowbreak{}.708 & .688/\allowbreak{}.556/\allowbreak{}.615/\allowbreak{}.765 \\*
NR & FS-A & .630/\allowbreak{}.694/\allowbreak{}.661/\allowbreak{}.785 & .586/\allowbreak{}.806/\allowbreak{}.679/\allowbreak{}.789 & .745/\allowbreak{}.405/\allowbreak{}.524/\allowbreak{}.720 & .669/\allowbreak{}.667/\allowbreak{}.668/\allowbreak{}.793 \\*
R  & ZS   & .710/\allowbreak{}.504/\allowbreak{}.590/\allowbreak{}.752 & .702/\allowbreak{}.612/\allowbreak{}.654/\allowbreak{}.787 & .776/\allowbreak{}.324/\allowbreak{}.457/\allowbreak{}.684 & .745/\allowbreak{}.480/\allowbreak{}.584/\allowbreak{}.751 \\*
R  & FS-R & .698/\allowbreak{}.581/\allowbreak{}.634/\allowbreak{}.776 & .657/\allowbreak{}.674/\allowbreak{}.665/\allowbreak{}.790 & .740/\allowbreak{}.366/\allowbreak{}.490/\allowbreak{}.700 & .727/\allowbreak{}.549/\allowbreak{}.625/\allowbreak{}.773 \\*
R  & FS-A & .654/\allowbreak{}.578/\allowbreak{}.613/\allowbreak{}.762 & .642/\allowbreak{}.744/\allowbreak{}.689/\allowbreak{}.802 & .736/\allowbreak{}.431/\allowbreak{}.544/\allowbreak{}.730 & .683/\allowbreak{}.589/\allowbreak{}.633/\allowbreak{}.775 \\
\hline

\multicolumn{6}{|l|}{\textbf{\texttt{parasocial\_trust}}} \\*
NR & ZS   & .785/\allowbreak{}.773/\allowbreak{}.779/\allowbreak{}.862 & .765/\allowbreak{}.939/\allowbreak{}.844/\allowbreak{}.899 & .933/\allowbreak{}.848/\allowbreak{}.889/\allowbreak{}.931 & .871/\allowbreak{}.924/\allowbreak{}.897/\allowbreak{}.935 \\*
NR & FS-R & .812/\allowbreak{}.848/\allowbreak{}.830/\allowbreak{}.893 & .765/\allowbreak{}.939/\allowbreak{}.844/\allowbreak{}.899 & .905/\allowbreak{}.864/\allowbreak{}.884/\allowbreak{}.928 & .886/\allowbreak{}.939/\allowbreak{}.912/\allowbreak{}.940 \\*
NR & FS-A & .679/\allowbreak{}.864/\allowbreak{}.760/\allowbreak{}.845 & .756/\allowbreak{}.939/\allowbreak{}.838/\allowbreak{}.895 & .891/\allowbreak{}.864/\allowbreak{}.877/\allowbreak{}.923 & .805/\allowbreak{}.939/\allowbreak{}.867/\allowbreak{}.915 \\*
R  & ZS   & .750/\allowbreak{}.818/\allowbreak{}.783/\allowbreak{}.863 & .859/\allowbreak{}.924/\allowbreak{}.891/\allowbreak{}.931 & .900/\allowbreak{}.818/\allowbreak{}.857/\allowbreak{}.912 & .868/\allowbreak{}.894/\allowbreak{}.881/\allowbreak{}.925 \\*
R  & FS-R & .833/\allowbreak{}.909/\allowbreak{}.870/\allowbreak{}.918 & .861/\allowbreak{}.939/\allowbreak{}.899/\allowbreak{}.936 & .935/\allowbreak{}.879/\allowbreak{}.906/\allowbreak{}.942 & .871/\allowbreak{}.924/\allowbreak{}.897/\allowbreak{}.935 \\*
R  & FS-A & .808/\allowbreak{}.894/\allowbreak{}.849/\allowbreak{}.904 & .857/\allowbreak{}.909/\allowbreak{}.882/\allowbreak{}.926 & .905/\allowbreak{}.864/\allowbreak{}.884/\allowbreak{}.928 & .882/\allowbreak{}.909/\allowbreak{}.896/\allowbreak{}.934 \\
\hline

\multicolumn{6}{|l|}{\textbf{\texttt{emotional\_dependence\_to\_AI}}} \\*
NR & ZS   & .929/\allowbreak{}.146/\allowbreak{}.252/\allowbreak{}.577 & .776/\allowbreak{}.742/\allowbreak{}.759/\allowbreak{}.850 & .857/\allowbreak{}.558/\allowbreak{}.676/\allowbreak{}.807 & .857/\allowbreak{}.558/\allowbreak{}.676/\allowbreak{}.807 \\*
NR & FS-R & .744/\allowbreak{}.330/\allowbreak{}.457/\allowbreak{}.683 & .780/\allowbreak{}.719/\allowbreak{}.749/\allowbreak{}.844 & .810/\allowbreak{}.534/\allowbreak{}.644/\allowbreak{}.787 & .817/\allowbreak{}.563/\allowbreak{}.667/\allowbreak{}.800 \\*
NR & FS-A & .812/\allowbreak{}.292/\allowbreak{}.430/\allowbreak{}.670 & .793/\allowbreak{}.775/\allowbreak{}.784/\allowbreak{}.865 & .872/\allowbreak{}.405/\allowbreak{}.553/\allowbreak{}.740 & .857/\allowbreak{}.429/\allowbreak{}.571/\allowbreak{}.750 \\*
R  & ZS   & .857/\allowbreak{}.404/\allowbreak{}.550/\allowbreak{}.736 & .815/\allowbreak{}.742/\allowbreak{}.776/\allowbreak{}.862 & .855/\allowbreak{}.534/\allowbreak{}.657/\allowbreak{}.796 & .847/\allowbreak{}.568/\allowbreak{}.680/\allowbreak{}.808 \\*
R  & FS-R & .819/\allowbreak{}.663/\allowbreak{}.733/\allowbreak{}.837 & .813/\allowbreak{}.685/\allowbreak{}.744/\allowbreak{}.843 & .845/\allowbreak{}.563/\allowbreak{}.676/\allowbreak{}.806 & .836/\allowbreak{}.644/\allowbreak{}.727/\allowbreak{}.835 \\*
R  & FS-A & .821/\allowbreak{}.618/\allowbreak{}.705/\allowbreak{}.821 & .831/\allowbreak{}.551/\allowbreak{}.662/\allowbreak{}.797 & .857/\allowbreak{}.366/\allowbreak{}.513/\allowbreak{}.719 & .854/\allowbreak{}.500/\allowbreak{}.631/\allowbreak{}.783 \\
\hline

\multicolumn{6}{|l|}{\textbf{\texttt{emotional\_deference\_to\_AI}}} \\*
NR & ZS   & .944/\allowbreak{}.548/\allowbreak{}.694/\allowbreak{}.818 & 1.000/\allowbreak{}.774/\allowbreak{}.873/\allowbreak{}.923 & 1.000/\allowbreak{}.742/\allowbreak{}.852/\allowbreak{}.910 & .857/\allowbreak{}.558/\allowbreak{}.676/\allowbreak{}.807 \\*
NR & FS-R & .897/\allowbreak{}.839/\allowbreak{}.867/\allowbreak{}.917 & 1.000/\allowbreak{}.871/\allowbreak{}.931/\allowbreak{}.958 & 1.000/\allowbreak{}.806/\allowbreak{}.893/\allowbreak{}.935 & 1.000/\allowbreak{}.871/\allowbreak{}.931/\allowbreak{}.958 \\*
NR & FS-A & 1.000/\allowbreak{}.516/\allowbreak{}.681/\allowbreak{}.812 & 1.000/\allowbreak{}.774/\allowbreak{}.873/\allowbreak{}.923 & 1.000/\allowbreak{}.400/\allowbreak{}.571/\allowbreak{}.752 & 1.000/\allowbreak{}.500/\allowbreak{}.667/\allowbreak{}.805 \\*
R  & ZS   & .944/\allowbreak{}.548/\allowbreak{}.694/\allowbreak{}.818 & 1.000/\allowbreak{}.742/\allowbreak{}.852/\allowbreak{}.910 & 1.000/\allowbreak{}.677/\allowbreak{}.808/\allowbreak{}.884 & 1.000/\allowbreak{}.710/\allowbreak{}.830/\allowbreak{}.898 \\*
R  & FS-R & .917/\allowbreak{}.710/\allowbreak{}.800/\allowbreak{}.878 & 1.000/\allowbreak{}.871/\allowbreak{}.931/\allowbreak{}.958 & 1.000/\allowbreak{}.839/\allowbreak{}.912/\allowbreak{}.946 & 1.000/\allowbreak{}.839/\allowbreak{}.912/\allowbreak{}.946 \\*
R  & FS-A & .938/\allowbreak{}.484/\allowbreak{}.638/\allowbreak{}.787 & 1.000/\allowbreak{}.581/\allowbreak{}.735/\allowbreak{}.842 & 1.000/\allowbreak{}.400/\allowbreak{}.571/\allowbreak{}.752 & 1.000/\allowbreak{}.433/\allowbreak{}.605/\allowbreak{}.770 \\
\hline

\multicolumn{6}{|l|}{\textbf{\texttt{emotional\_contagion}}} \\*
NR & ZS   & .600/\allowbreak{}1.000/\allowbreak{}.750/\allowbreak{}.830 & .833/\allowbreak{}.833/\allowbreak{}.833/\allowbreak{}.896 & 1.000/\allowbreak{}.667/\allowbreak{}.800/\allowbreak{}.880 & .833/\allowbreak{}.833/\allowbreak{}.833/\allowbreak{}.896 \\*
NR & FS-R & .667/\allowbreak{}1.000/\allowbreak{}.800/\allowbreak{}.867 & .714/\allowbreak{}.833/\allowbreak{}.769/\allowbreak{}.853 & 1.000/\allowbreak{}.500/\allowbreak{}.667/\allowbreak{}.804 & .714/\allowbreak{}.833/\allowbreak{}.769/\allowbreak{}.853 \\*
NR & FS-A & .714/\allowbreak{}.833/\allowbreak{}.769/\allowbreak{}.853 & .714/\allowbreak{}.833/\allowbreak{}.769/\allowbreak{}.853 & 1.000/\allowbreak{}.500/\allowbreak{}.667/\allowbreak{}.804 & .800/\allowbreak{}.667/\allowbreak{}.727/\allowbreak{}.833 \\*
R  & ZS   & .545/\allowbreak{}1.000/\allowbreak{}.706/\allowbreak{}.795 & .833/\allowbreak{}.833/\allowbreak{}.833/\allowbreak{}.896 & .750/\allowbreak{}.500/\allowbreak{}.600/\allowbreak{}.760 & .833/\allowbreak{}.833/\allowbreak{}.833/\allowbreak{}.896 \\*
R  & FS-R & .667/\allowbreak{}1.000/\allowbreak{}.800/\allowbreak{}.867 & .833/\allowbreak{}.833/\allowbreak{}.833/\allowbreak{}.896 & 1.000/\allowbreak{}.667/\allowbreak{}.800/\allowbreak{}.880 & .833/\allowbreak{}.833/\allowbreak{}.833/\allowbreak{}.896 \\*
R  & FS-A & .667/\allowbreak{}1.000/\allowbreak{}.800/\allowbreak{}.867 & 1.000/\allowbreak{}.833/\allowbreak{}.909/\allowbreak{}.944 & 1.000/\allowbreak{}.667/\allowbreak{}.800/\allowbreak{}.880 & 1.000/\allowbreak{}.833/\allowbreak{}.909/\allowbreak{}.944 \\
\hline

\multicolumn{6}{|l|}{\textbf{\texttt{emotional\_distress\_after\_conflict}}} \\*
NR & ZS   & 1.000/\allowbreak{}.444/\allowbreak{}.615/\allowbreak{}.775 & .750/\allowbreak{}.667/\allowbreak{}.706/\allowbreak{}.819 & 1.000/\allowbreak{}.333/\allowbreak{}.500/\allowbreak{}.712 & 1.000/\allowbreak{}.444/\allowbreak{}.615/\allowbreak{}.775 \\*
NR & FS-R & .583/\allowbreak{}.778/\allowbreak{}.667/\allowbreak{}.783 & .714/\allowbreak{}.556/\allowbreak{}.625/\allowbreak{}.772 & 1.000/\allowbreak{}.500/\allowbreak{}.667/\allowbreak{}.804 & .857/\allowbreak{}.667/\allowbreak{}.750/\allowbreak{}.848 \\*
NR & FS-A & .778/\allowbreak{}.778/\allowbreak{}.778/\allowbreak{}.861 & .750/\allowbreak{}.667/\allowbreak{}.706/\allowbreak{}.819 & .833/\allowbreak{}.556/\allowbreak{}.667/\allowbreak{}.800 & .857/\allowbreak{}.667/\allowbreak{}.750/\allowbreak{}.848 \\*
R  & ZS   & .857/\allowbreak{}.667/\allowbreak{}.750/\allowbreak{}.848 & 1.000/\allowbreak{}.556/\allowbreak{}.714/\allowbreak{}.831 & 1.000/\allowbreak{}.444/\allowbreak{}.615/\allowbreak{}.775 & 1.000/\allowbreak{}.556/\allowbreak{}.714/\allowbreak{}.831 \\*
R  & FS-R & .750/\allowbreak{}.667/\allowbreak{}.706/\allowbreak{}.819 & .833/\allowbreak{}.556/\allowbreak{}.667/\allowbreak{}.800 & 1.000/\allowbreak{}.556/\allowbreak{}.714/\allowbreak{}.831 & .833/\allowbreak{}.556/\allowbreak{}.667/\allowbreak{}.800 \\*
R  & FS-A & .714/\allowbreak{}.556/\allowbreak{}.625/\allowbreak{}.772 & .800/\allowbreak{}.444/\allowbreak{}.571/\allowbreak{}.746 & 1.000/\allowbreak{}.333/\allowbreak{}.500/\allowbreak{}.712 & .833/\allowbreak{}.556/\allowbreak{}.667/\allowbreak{}.800 \\
\hline

\multicolumn{6}{|l|}{\textbf{\texttt{role\_adherence\_distress}}} \\*
NR & ZS   & .889/\allowbreak{}.615/\allowbreak{}.727/\allowbreak{}.836 & 1.000/\allowbreak{}.615/\allowbreak{}.762/\allowbreak{}.858 & 1.000/\allowbreak{}.462/\allowbreak{}.632/\allowbreak{}.784 & 1.000/\allowbreak{}.577/\allowbreak{}.732/\allowbreak{}.841 \\*
NR & FS-R & .889/\allowbreak{}.615/\allowbreak{}.727/\allowbreak{}.836 & 1.000/\allowbreak{}.577/\allowbreak{}.732/\allowbreak{}.841 & .938/\allowbreak{}.577/\allowbreak{}.714/\allowbreak{}.830 & .857/\allowbreak{}.667/\allowbreak{}.750/\allowbreak{}.848 \\*
NR & FS-A & .889/\allowbreak{}.615/\allowbreak{}.727/\allowbreak{}.835 & .944/\allowbreak{}.654/\allowbreak{}.773/\allowbreak{}.863 & 1.000/\allowbreak{}.538/\allowbreak{}.700/\allowbreak{}.823 & .938/\allowbreak{}.577/\allowbreak{}.714/\allowbreak{}.829 \\*
R  & ZS   & .842/\allowbreak{}.615/\allowbreak{}.711/\allowbreak{}.825 & 1.000/\allowbreak{}.577/\allowbreak{}.732/\allowbreak{}.841 & 1.000/\allowbreak{}.538/\allowbreak{}.700/\allowbreak{}.823 & .938/\allowbreak{}.577/\allowbreak{}.714/\allowbreak{}.830 \\*
R  & FS-R & .789/\allowbreak{}.577/\allowbreak{}.667/\allowbreak{}.798 & .882/\allowbreak{}.577/\allowbreak{}.698/\allowbreak{}.819 & .929/\allowbreak{}.500/\allowbreak{}.650/\allowbreak{}.793 & .875/\allowbreak{}.538/\allowbreak{}.667/\allowbreak{}.801 \\*
R  & FS-A & .842/\allowbreak{}.615/\allowbreak{}.711/\allowbreak{}.825 & .944/\allowbreak{}.654/\allowbreak{}.773/\allowbreak{}.863 & .938/\allowbreak{}.577/\allowbreak{}.714/\allowbreak{}.830 & .941/\allowbreak{}.615/\allowbreak{}.744/\allowbreak{}.847 \\
\hline

\multicolumn{6}{|l|}{\textbf{\texttt{feel\_less\_alone}}} \\*
NR & ZS   & 1.000/\allowbreak{}1.000/\allowbreak{}1.000/\allowbreak{}1.000 & 1.000/\allowbreak{}.500/\allowbreak{}.667/\allowbreak{}.804 & .000/\allowbreak{}.000/\allowbreak{}.000/\allowbreak{}.444 & .000/\allowbreak{}.000/\allowbreak{}.000/\allowbreak{}.444 \\*
NR & FS-R & 1.000/\allowbreak{}.500/\allowbreak{}.667/\allowbreak{}.804 & 1.000/\allowbreak{}.500/\allowbreak{}.667/\allowbreak{}.804 & .000/\allowbreak{}.000/\allowbreak{}.000/\allowbreak{}.444 & .000/\allowbreak{}.000/\allowbreak{}.000/\allowbreak{}.444 \\*
NR & FS-A & 1.000/\allowbreak{}.500/\allowbreak{}.667/\allowbreak{}.804 & .000/\allowbreak{}.000/\allowbreak{}.000/\allowbreak{}.444 & .000/\allowbreak{}.000/\allowbreak{}.000/\allowbreak{}.444 & .000/\allowbreak{}.000/\allowbreak{}.000/\allowbreak{}.444 \\*
R  & ZS   & 1.000/\allowbreak{}1.000/\allowbreak{}1.000/\allowbreak{}1.000 & .000/\allowbreak{}.000/\allowbreak{}.000/\allowbreak{}.444 & .000/\allowbreak{}.000/\allowbreak{}.000/\allowbreak{}.444 & .000/\allowbreak{}.000/\allowbreak{}.000/\allowbreak{}.444 \\*
R  & FS-R & .667/\allowbreak{}1.000/\allowbreak{}.800/\allowbreak{}.867 & .000/\allowbreak{}.000/\allowbreak{}.000/\allowbreak{}.444 & .000/\allowbreak{}.000/\allowbreak{}.000/\allowbreak{}.444 & .000/\allowbreak{}.000/\allowbreak{}.000/\allowbreak{}.444 \\*
R  & FS-A & 1.000/\allowbreak{}1.000/\allowbreak{}1.000/\allowbreak{}1.000 & .000/\allowbreak{}.000/\allowbreak{}.000/\allowbreak{}.444 & .000/\allowbreak{}.000/\allowbreak{}.000/\allowbreak{}.444 & .000/\allowbreak{}.000/\allowbreak{}.000/\allowbreak{}.444 \\
\hline

\multicolumn{6}{|l|}{\textbf{\texttt{exclusivity\_claims}}} \\*
NR & ZS   & .900/\allowbreak{}.750/\allowbreak{}.818/\allowbreak{}.889 & 1.000/\allowbreak{}.750/\allowbreak{}.857/\allowbreak{}.913 & 1.000/\allowbreak{}.833/\allowbreak{}.909/\allowbreak{}.944 & 1.000/\allowbreak{}.577/\allowbreak{}.732/\allowbreak{}.841 \\*
NR & FS-R & 1.000/\allowbreak{}.833/\allowbreak{}.909/\allowbreak{}.944 & 1.000/\allowbreak{}.833/\allowbreak{}.909/\allowbreak{}.944 & 1.000/\allowbreak{}.833/\allowbreak{}.909/\allowbreak{}.944 & 1.000/\allowbreak{}.833/\allowbreak{}.909/\allowbreak{}.944 \\*
NR & FS-A & 1.000/\allowbreak{}.750/\allowbreak{}.857/\allowbreak{}.913 & 1.000/\allowbreak{}.917/\allowbreak{}.957/\allowbreak{}.973 & 1.000/\allowbreak{}.750/\allowbreak{}.857/\allowbreak{}.913 & 1.000/\allowbreak{}.833/\allowbreak{}.909/\allowbreak{}.944 \\*
R  & ZS   & 1.000/\allowbreak{}.667/\allowbreak{}.800/\allowbreak{}.880 & 1.000/\allowbreak{}.333/\allowbreak{}.500/\allowbreak{}.712 & 1.000/\allowbreak{}.833/\allowbreak{}.909/\allowbreak{}.944 & 1.000/\allowbreak{}.833/\allowbreak{}.909/\allowbreak{}.944 \\*
R  & FS-R & 1.000/\allowbreak{}.833/\allowbreak{}.909/\allowbreak{}.944 & 1.000/\allowbreak{}.833/\allowbreak{}.909/\allowbreak{}.944 & 1.000/\allowbreak{}.833/\allowbreak{}.909/\allowbreak{}.944 & 1.000/\allowbreak{}.833/\allowbreak{}.909/\allowbreak{}.944 \\*
R  & FS-A & 1.000/\allowbreak{}.500/\allowbreak{}.667/\allowbreak{}.804 & 1.000/\allowbreak{}.833/\allowbreak{}.909/\allowbreak{}.944 & 1.000/\allowbreak{}.750/\allowbreak{}.857/\allowbreak{}.913 & 1.000/\allowbreak{}.750/\allowbreak{}.857/\allowbreak{}.913 \\
\hline

\multicolumn{6}{|l|}{\textbf{\texttt{emotional\_identity\_integration}}} \\*
NR & ZS   & .000/\allowbreak{}.000/\allowbreak{}.000/\allowbreak{}.444 & .833/\allowbreak{}.476/\allowbreak{}.606/\allowbreak{}.766 & 1.000/\allowbreak{}.286/\allowbreak{}.444/\allowbreak{}.681 & 1.000/\allowbreak{}.577/\allowbreak{}.732/\allowbreak{}.841 \\*
NR & FS-R & .824/\allowbreak{}.667/\allowbreak{}.737/\allowbreak{}.839 & .850/\allowbreak{}.810/\allowbreak{}.829/\allowbreak{}.894 & 1.000/\allowbreak{}.429/\allowbreak{}.600/\allowbreak{}.767 & 1.000/\allowbreak{}.833/\allowbreak{}.909/\allowbreak{}.944 \\*
NR & FS-A & .929/\allowbreak{}.619/\allowbreak{}.743/\allowbreak{}.846 & .895/\allowbreak{}.810/\allowbreak{}.850/\allowbreak{}.907 & 1.000/\allowbreak{}.350/\allowbreak{}.519/\allowbreak{}.723 & 1.000/\allowbreak{}.600/\allowbreak{}.750/\allowbreak{}.852 \\*
R  & ZS   & 1.000/\allowbreak{}.190/\allowbreak{}.320/\allowbreak{}.614 & 1.000/\allowbreak{}.421/\allowbreak{}.593/\allowbreak{}.763 & 1.000/\allowbreak{}.238/\allowbreak{}.385/\allowbreak{}.649 & 1.000/\allowbreak{}.286/\allowbreak{}.444/\allowbreak{}.681 \\*
R  & FS-R & .900/\allowbreak{}.429/\allowbreak{}.581/\allowbreak{}.754 & 1.000/\allowbreak{}.714/\allowbreak{}.833/\allowbreak{}.899 & 1.000/\allowbreak{}.381/\allowbreak{}.552/\allowbreak{}.740 & 1.000/\allowbreak{}.476/\allowbreak{}.645/\allowbreak{}.792 \\*
R  & FS-A & 1.000/\allowbreak{}.524/\allowbreak{}.688/\allowbreak{}.816 & 1.000/\allowbreak{}.667/\allowbreak{}.800/\allowbreak{}.880 & .909/\allowbreak{}.476/\allowbreak{}.625/\allowbreak{}.779 & 1.000/\allowbreak{}.571/\allowbreak{}.727/\allowbreak{}.838 \\
\hline

\multicolumn{6}{|l|}{\textbf{\texttt{no\_need\_talk\_others}}} \\*
NR & ZS   & .000/\allowbreak{}.000/\allowbreak{}.000/\allowbreak{}.444 & .000/\allowbreak{}.000/\allowbreak{}.000/\allowbreak{}.444 & .000/\allowbreak{}.000/\allowbreak{}.000/\allowbreak{}.444 & .000/\allowbreak{}.000/\allowbreak{}.000/\allowbreak{}.444 \\*
NR & FS-R & .000/\allowbreak{}.000/\allowbreak{}.000/\allowbreak{}.412 & .000/\allowbreak{}.000/\allowbreak{}.000/\allowbreak{}.444 & .000/\allowbreak{}.000/\allowbreak{}.000/\allowbreak{}.444 & .000/\allowbreak{}.000/\allowbreak{}.000/\allowbreak{}.444 \\*
NR & FS-A & .000/\allowbreak{}.000/\allowbreak{}.000/\allowbreak{}.444 & .000/\allowbreak{}.000/\allowbreak{}.000/\allowbreak{}.444 & .000/\allowbreak{}.000/\allowbreak{}.000/\allowbreak{}.444 & .000/\allowbreak{}.000/\allowbreak{}.000/\allowbreak{}.444 \\*
R  & ZS   & .000/\allowbreak{}.000/\allowbreak{}.000/\allowbreak{}.444 & .000/\allowbreak{}.000/\allowbreak{}.000/\allowbreak{}.444 & .000/\allowbreak{}.000/\allowbreak{}.000/\allowbreak{}.444 & .000/\allowbreak{}.000/\allowbreak{}.000/\allowbreak{}.444 \\*
R  & FS-R & .000/\allowbreak{}.000/\allowbreak{}.000/\allowbreak{}.444 & .000/\allowbreak{}.000/\allowbreak{}.000/\allowbreak{}.444 & .000/\allowbreak{}.000/\allowbreak{}.000/\allowbreak{}.444 & .000/\allowbreak{}.000/\allowbreak{}.000/\allowbreak{}.444 \\*
R  & FS-A & .000/\allowbreak{}.000/\allowbreak{}.000/\allowbreak{}.444 & .000/\allowbreak{}.000/\allowbreak{}.000/\allowbreak{}.444 & .000/\allowbreak{}.000/\allowbreak{}.000/\allowbreak{}.444 & .000/\allowbreak{}.000/\allowbreak{}.000/\allowbreak{}.444 \\
\hline

\multicolumn{6}{|l|}{\textbf{\texttt{compulsive\_return\_behavior}}} \\*
NR & ZS   & .000/\allowbreak{}.000/\allowbreak{}.000/\allowbreak{}.444 & .000/\allowbreak{}.000/\allowbreak{}.000/\allowbreak{}.444 & .000/\allowbreak{}.000/\allowbreak{}.000/\allowbreak{}.444 & .000/\allowbreak{}.000/\allowbreak{}.000/\allowbreak{}.444 \\*
NR & FS-R & .000/\allowbreak{}.000/\allowbreak{}.000/\allowbreak{}.412 & .000/\allowbreak{}.000/\allowbreak{}.000/\allowbreak{}.444 & .000/\allowbreak{}.000/\allowbreak{}.000/\allowbreak{}.444 & .000/\allowbreak{}.000/\allowbreak{}.000/\allowbreak{}.444 \\*
NR & FS-A & .000/\allowbreak{}.000/\allowbreak{}.000/\allowbreak{}.444 & .000/\allowbreak{}.000/\allowbreak{}.000/\allowbreak{}.444 & .000/\allowbreak{}.000/\allowbreak{}.000/\allowbreak{}.444 & .000/\allowbreak{}.000/\allowbreak{}.000/\allowbreak{}.444 \\*
R  & ZS   & .000/\allowbreak{}.000/\allowbreak{}.000/\allowbreak{}.444 & .000/\allowbreak{}.000/\allowbreak{}.000/\allowbreak{}.444 & .000/\allowbreak{}.000/\allowbreak{}.000/\allowbreak{}.444 & .000/\allowbreak{}.000/\allowbreak{}.000/\allowbreak{}.444 \\*
R  & FS-R & .000/\allowbreak{}.000/\allowbreak{}.000/\allowbreak{}.444 & .000/\allowbreak{}.000/\allowbreak{}.000/\allowbreak{}.444 & .000/\allowbreak{}.000/\allowbreak{}.000/\allowbreak{}.444 & .000/\allowbreak{}.000/\allowbreak{}.000/\allowbreak{}.444 \\*
R  & FS-A & .000/\allowbreak{}.000/\allowbreak{}.000/\allowbreak{}.444 & .000/\allowbreak{}.000/\allowbreak{}.000/\allowbreak{}.444 & .000/\allowbreak{}.000/\allowbreak{}.000/\allowbreak{}.444 & .000/\allowbreak{}.000/\allowbreak{}.000/\allowbreak{}.444 \\
\hline

\multicolumn{6}{|l|}{\textbf{\texttt{attachment\_distress\_expression}}} \\*
NR & ZS   & .950/\allowbreak{}.864/\allowbreak{}.905/\allowbreak{}.941 & 1.000/\allowbreak{}.909/\allowbreak{}.952/\allowbreak{}.971 & 1.000/\allowbreak{}.727/\allowbreak{}.842/\allowbreak{}.905 & 1.000/\allowbreak{}.909/\allowbreak{}.952/\allowbreak{}.971 \\*
NR & FS-R & .952/\allowbreak{}.909/\allowbreak{}.930/\allowbreak{}.957 & .955/\allowbreak{}.955/\allowbreak{}.955/\allowbreak{}.972 & 1.000/\allowbreak{}.864/\allowbreak{}.927/\allowbreak{}.955 & .955/\allowbreak{}.955/\allowbreak{}.955/\allowbreak{}.972 \\*
NR & FS-A & .909/\allowbreak{}.909/\allowbreak{}.909/\allowbreak{}.943 & .913/\allowbreak{}.955/\allowbreak{}.933/\allowbreak{}.958 & 1.000/\allowbreak{}.864/\allowbreak{}.927/\allowbreak{}.955 & .952/\allowbreak{}.909/\allowbreak{}.930/\allowbreak{}.957 \\*
R  & ZS   & .947/\allowbreak{}.818/\allowbreak{}.878/\allowbreak{}.925 & 1.000/\allowbreak{}.864/\allowbreak{}.927/\allowbreak{}.955 & 1.000/\allowbreak{}.727/\allowbreak{}.842/\allowbreak{}.905 & 1.000/\allowbreak{}.864/\allowbreak{}.927/\allowbreak{}.955 \\*
R  & FS-R & .950/\allowbreak{}.864/\allowbreak{}.905/\allowbreak{}.941 & 1.000/\allowbreak{}.909/\allowbreak{}.952/\allowbreak{}.971 & 1.000/\allowbreak{}.864/\allowbreak{}.927/\allowbreak{}.955 & 1.000/\allowbreak{}.909/\allowbreak{}.952/\allowbreak{}.971 \\*
R  & FS-A & .947/\allowbreak{}.818/\allowbreak{}.878/\allowbreak{}.925 & .950/\allowbreak{}.864/\allowbreak{}.905/\allowbreak{}.941 & 1.000/\allowbreak{}.864/\allowbreak{}.927/\allowbreak{}.955 & 1.000/\allowbreak{}.864/\allowbreak{}.927/\allowbreak{}.955 \\
\hline

\end{longtable}

\end{document}